\documentclass[preprint]{aastex}
\usepackage{amsmath}
\begin{document}

\title{ALMA Observations of Infalling Flows toward the Keplerian Disk around the Class I Protostar L1489 IRS}
\author{Hsi-Wei Yen\altaffilmark{1}, Shigehisa Takakuwa\altaffilmark{1}, Nagayoshi Ohashi\altaffilmark{1,2}, Yuri Aikawa\altaffilmark{3}, Yusuke Aso\altaffilmark{4}, Shin Koyamatsu\altaffilmark{4,2}, Masahiro N. Machida\altaffilmark{5}, Kazuya Saigo\altaffilmark{6}, Masao Saito\altaffilmark{7}, Kengo Tomida\altaffilmark{8,9}, and Kohji Tomisaka\altaffilmark{10,11}}









\altaffiltext{1}{Academia Sinica Institute of Astronomy and Astrophysics, P.O. Box 23-141, Taipei 10617, Taiwan; hwyen@asiaa.sinica.edu.tw} 
\altaffiltext{2}{Subaru Telescope, National Astronomical Observatory of Japan, 650 North A'ohoku Place, Hilo, HI 96720, USA}
\altaffiltext{3}{Department of Earth and Planetary Sciences, Kobe University, Kobe 657-8501, Japan}
\altaffiltext{4}{Department of Astronomy, Graduate School of Science, The University of Tokyo, 7-3-1 Hongo, Bunkyo-ku, Tokyo 113-0033, Japan}
\altaffiltext{5}{Department of Earth and Planetary Sciences, Faculty of Sciences, Kyushu University, Fukuoka 812-8581, Japan}
\altaffiltext{6}{Chile Observatory, National Astronomical Observatory of Japan, Osawa 2-21-1, Mitaka, Tokyo 181-8588, Japan}
\altaffiltext{7}{Joint ALMA Observatory, Ave. Alonso de Cordova 3107, Vitacura, Santiago, Chile}
\altaffiltext{8}{Department of Astrophysical Sciences, Princeton University, Princeton, NJ 08544, USA}
\altaffiltext{9}{Department of Physics, The University of Tokyo, Tokyo 113-0033, Japan}
\altaffiltext{10}{Department of Astronomical Science, The Graduate University for Advanced Studies (SOKENDAI), Osawa, Mitaka, Tokyo 181-8588, Japan} 
\altaffiltext{11}{National Astronomical Observatory of Japan, Osawa, Mitaka, Tokyo 181-8588, Japan}

\begin{abstract}
We have conducted ALMA observations in the 1.3 mm continuum and $^{12}$CO (2--1), C$^{18}$O (2--1) and SO (5$_6$--4$_5$) lines toward L1489 IRS, a Class I protostar surrounded by a Keplerian disk and an infalling envelope. 
The Keplerian disk is clearly identified in the $^{12}$CO and C$^{18}$O emission, 
and its outer radius ($\sim$700 AU) and mass ($\sim$0.005 $M_\sun$) are comparable to those of disks around T Tauri stars. 
The protostellar mass is estimated to be 1.6 $M_\sun$ with the inclination angle of 66\degr. 
In addition to the Keplerian disk, 
there are blueshifted and redshifted off-axis protrusions seen in the C$^{18}$O emission pointing toward the north and the south, respectively, adjunct to the middle part of the Keplerian disk.  
The shape and kinematics of these protrusions can be interpreted as streams of infalling flows with a conserved angular momentum following parabolic trajectories toward the Keplerian disk, 
and the mass infalling rate is estimated to be $\sim$5 $\times$ 10$^{-7}$ $M_\sun$ yr$^{-1}$. 
The specific angular momentum of the infalling flows ($\sim$2.5 $\times$ 10$^{-3}$ km s$^{-1}$ pc) is comparable to that at the outer radius of the Keplerian disk ($\sim$4.8 $\times$ 10$^{-3}$ km s$^{-1}$ pc).
The SO emission is elongated along the disk major axis and exhibits a linear velocity gradient along the axis, 
which is interpreted as that the SO emission primarily traces a ring region in the flared Keplerian disk at radii of $\sim$250--390 AU. 
The local enhancement of the SO abundance in the ring region can be due to the accretion shocks at the centrifugal radius where the infalling flows fall onto the disk.  
Our ALMA observations unveiled both the Keplerian disk and the infalling gas onto the disk, 
and the disk can further grow by accreting material and angular momenta from the infalling gas.

  
\end{abstract}

\section{Introduction}
Protostars form through gravitational collapse of dense cores ($n \sim 10^{4} - 10^{5}$ cm$^{-3}$) in molecular clouds (e.g., Shu et al.~1987; Andr\'{e} et al.~2000; Myers et al.~2000).     
Previous interferometric observations have found infalling and rotational motions on thousands of AU scale inside dense cores associated with known protostellar sources, so-called protostellar envelopes (e.g., Ohashi et al.~1997; Momose et al.~1998). 
Keplerian disks have often been observed around T Tauri stars with (sub)millimeter interferometric observations (e.g., Guilloteau \& Dutrey 1994; Dutrey et al.~1998; Simon et al.~2000; Qi et al.~2003, 2004; Andrews \& Williams 2007; Pietu et al. 2007; Guilloteau et al.~2011; Andrews et al.~2012; P{\'e}rez et al.~2012). 
The outer radii of those Keplerian disks seen in CO emission range from $\sim$100 AU to $\sim$800 AU, 
and their masses traced by dust continuum emission range from $\sim$$10^{-4}$ $M_\sun$ to $\sim$10$^{-1}$ $M_\sun$. 
Recent interferometric observations have detected Keplerian disks around Class I protostars, which are still embedded in protostellar envelopes (e.g., Brinch et al.~2007b; Lommen et al.~2008; J{\o}rgensen et al.~2009; Takakuwa et al.~2012; Yen et al.~2013; Harsono et al.~2014). 
The Keplerian disks around Class I protostars have outer radii ranging from 80 AU to 300 AU and masses from 0.004 $M_\sun$ to 0.08 $M_\sun$, comparable to those around T Tauri stars. 
Recently, Keplerian disks have also been observed around the Class 0 protostar, VLA 1623 (Murillo et al.~2013), and the protostars at the transitional stage from Class 0 to I, L1527 IRS (Tobin et al.~2012a) and [BHB2007]\#11 (Hara et al.~2013). 
The outer radii of the Keplerian disks around these protostars at the earlier evolutionary stages range from 90 AU to 350 AU, comparable to those around Class I protostars.  
However, it is still controversial how these large-scale ($\sim$100 AU) Keplerian disks form in infalling and rotating protostellar envelopes. 

Theoretical models of a collapsing dense core without magnetic field suggest that a Keplerian disk form around a protostar when collapsing material rotates fast enough to become centrifugally supported, 
and that the outer radius of the Keplerian disk increases as the collapse proceeds toward outer regions (Ulrich 1976; Cassen \& Moosman 1981; Terebey et al.~1984; Basu 1998; Bate 1998). 
On the other hand, 
previous magnetohydrodynamic (MHD) simulations show that the magnetic field can effectively remove the angular momentum of collapsing material by magnetic braking, 
and suppress the outer radii of Keplerian disks within $\sim$10 AU (e.g., Mellon \& Li 2008, 2009; Machida et al.~2011; Li et al.~2011; Dapp et al.~2012; Tomida et al.~2013). 
In these MHD simulations, large-scale Keplerian disks cannot form until the efficiency of magnetic braking decreases due to the dissipation of protostellar envelopes (Machida et al.~2011; Machida \& Hosokawa 2013). 
Recent theoretical simulations by Machida et al.~(2014), investigating dependence of the initial model settings such as the sink radius and the density profile on the disk formation, have successfully reproduced Keplerian disks with similar properties to those of observed Keplerian disks around protostars. 

To study evolution from infalling and rotating envelopes to large-scale Keplerian disks observationally, 
we have conducted Submillimeter Array (SMA) observations of the inner part (100--1000 AU) of protostellar envelopes around a sample of Class 0 and I protostars (Yen et al.~2010, 2011, 2013). 
Our observations, coupled with previous observations of protostellar envelopes (e.g., Arce \& Sargent 2006), show that the kinematics of protostellar envelopes on 100--1000 AU scale can be categorized into three groups, (1) infalling motion with little rotational motion around Class 0 protostars (e.g., B335 and NGC 1333 IRAS 4B), (2) both infalling and rotational motions around Class 0 and I protostars (e.g., L1448-mm and L1527 IRS), and (3) Keplerian rotation around Class I protostars (e.g., TMC-1A and L1489 IRS). 
The three categories demonstrate the evolution from infalling envelopes to Keplerian disks, and can be explained by inside-out collapse of a rotating dense core where the angular momentum is conserved (Yen et al.~2013 and references therein). 
At the early collapse stage, the envelope material with a low angular momentum in the vicinity of a protostar collapses first, 
and hence the protostellar envelope shows infalling motion but little rotational motion. 
As the expansion wave propagates outward, the envelope material with a higher angular momentum in the outer region start to collapse and forms a Keplerian disk at the center.  
As more angular momenta travel to the central region with the infalling motion, rotational velocities of the envelope on 100--1000 AU scale increase, and the size of the central Keplerian disk expands.  
Eventually, the protostellar envelope is dissipated through the completion of the mass accretion onto the central protostar+disk system and/or the interaction with the outflow, 
and the central Keplerian disk with a radius of hundreds of AU emerges. 

Our next interest is direct imaging of the transition from an outer infalling envelope to an inner Keplerian disk in protostellar sources. 
Then, we should be able to discuss how outer infalling and rotational motions transform into Keplerian rotation and how an infalling envelope feeds material and angular momenta to a Keplerian disk, that is, the ongoing process of the disk growth. 
Takakuwa et al.~(2013) have found infalling gas outside of the Keplerian circumbinary disk around L1551 NE, 
and revealed that the angular momentum of the infalling gas is much lower than that of the Keplerian disk. 
The spatial resolution of their SMA observations is, however, not high enough to resolve the transitional region from the infalling gas to the Keplerian disk. 
Higher-resolution, higher-sensitivity observations of the transitional regions between the infalling and Keplerian regions are required to unveil the ongoing process of the disk formation.

We have conducted observations with the Atacama Large Millimeter/submillimeter Array (ALMA) in the 1.3 mm continuum emission and the $^{12}$CO (2--1; 230.538 GHz), C$^{18}$O (2--1; 219.560358 GHz), and SO (5$_6$--4$_5$; 219.949433 GHz) lines toward L1489 IRS, a Class I protostar with a bolometric luminosity of 3.7 $L_\sun$ (Furlan et al.~2008) in the Taurus molecular cloud ($d$ = 140 pc). 
L1489 IRS is associated with a faint molecular outflow along the north--south direction on thousands of AU scale as observed by the JCMT in the $^{12}$CO (3--2) line (Hogerheijde et al.~1998). 
Single-dish continuum observations at 1.3 mm show that L1489 IRS is surrounded by a compact protostellar envelope with a size of $\sim$30\arcsec\ ($\sim$4200 AU) and a mass of 0.03 $M_\sun$ (Motte \& Andr{\'e} 2001). 
Both infalling and rotational motions are present in the envelope on a 2000 AU scale, as found by interferometric observations in millimeter molecular lines, 
and the rotational motion is more dominant than the infalling motion (Hogerheijde 2001). 
The SMA observations in the millimeter continuum and HCO$^{+}$ (3--2) line emission have reported the presence of a Keplerian disk with a radius of 200 AU and a mass of 0.004 $M_\sun$ embedded in the envelope, 
and the protostellar mass is estimated to be 1.4 $M_\sun$ on the assumption that the inclination angle of the Keplerian disk is 40$\degr$ (Brinch et al.~2007b). 
The 1.3 mm continuum observations with the CARMA at a sub-arcsecond angular resolution ($\sim0\farcs8$) have shown that the disk radius is 250 -- 450 AU and the disk mass is 0.005 $M_\sun$ (Eisner 2012). 
The radial profile of the rotational motion on 100--1000 AU scale is $V_{\rm rot} \propto r^{-0.5\pm0.1}$ as measured by the SMA C$^{18}$O (2--1) observations, consistent with Keplerian rotation (Yen et al.~2013). 
From the observed Keplerian rotation, the protostellar mass is estimated to be $1.8\pm0.2$ $M_\sun$ on the assumption that the inclination angle of the Keplerian disk is 50$\degr$ (Yen et al.~2013). 
These observational results show that L1489 IRS is an excellent target to study the kinematics, structures, and connection from an infalling envelope to a Keplerian disk.

In the present paper, 
we report the imaging and analyses of our ALMA observations of L1489 IRS.
Although our previous SMA C$^{18}$O (2--1) observations at an angular resolution of $\sim$4\arcsec\ have shown that the overall gas motion on 100--1000 AU scale around L1489 IRS follows Keplerian rotation,  
the observed intensity distributions and kinematics cannot be fully explained by a single Keplerian disk. 
This suggests that there are likely non-Keplerian components in addition to the Keplerian disk, 
and that the spatial and velocity structures of the non-Keplerian components are not fully resolved by our SMA observations. 
With the ALMA observations, providing an angular resolution four times better and sensitivity more than ten times better than those of the previous SMA observations, 
the non-Keplerian gas components are clearly identified in the C$^{18}$O (2--1) and SO (5$_6$--4$_5$) emission, as well as the previously reported Keplerian disk in the $^{12}$CO (2--1) and C$^{18}$O (2--1) emission.   
We propose a model comprising a Keplerian disk and infalling flows toward the disk to explain the observed intensity distributions and velocity structures.  
The observational results are compared with our model and discussed in the context of the evolution of protostellar envelopes and Keplerian circumstellar disks.  

\section{Observations}
The ALMA Band 6 observations of L1489 IRS were made with its extended configuration of 23 antennas at the cycle 0 stage on 2012 August 9.
The pointing center is $\alpha$(J2000) = $4^{h}04^{m}42\fs85$, $\delta$(J2000) = $26\arcdeg18\arcmin56\farcs3$.  
L1489 IRS was observed for 74 minutes (on-source time). 
The projected $uv$ distances range from $\sim$13 m to $\sim$383 m, 
and our observations are insensitive to structures at the scales larger than 17$\arcsec$ ($\sim$2400 AU) at a 10\% level (Wilner \& Welch 1994). 
The correlator was configured in the Frequency Division Mode (FDM) to provide four independent spectral windows with a bandwidth of 234 MHz.
Each spectral window was assigned with 3840 spectral channels, 
and Hanning smoothing was applied to the spectral channels, 
resulting in a frequency resolution of 122 kHz. 
This frequency resolution corresponds to velocity resolutions of $\sim$0.17 km s$^{-1}$ for the C$^{18}$O and SO lines and $\sim$0.16 km s$^{-1}$ for the $^{12}$CO line. 
Emission-free channels in all of the four spectral windows were integrated to create the continuum image, 
which has a central frequency of 225 GHz and a bandwidth of $\sim$550 MHz. 
Calibration of the raw visibility data were performed with the standard reduction script for the Cycle 0 data, which uses tasks in Common Astronomy Software Applications (CASA). 
The absolute flux value of the amplitude calibrator, Callisto, was derived from the CASA Butler-JPL-Horizons 2010 model,
which gives an estimated systematic flux uncertainty of 10$\%$.
J0522$-$364 ($\sim$4.7 Jy) and J0325+226 ($\sim$0.3 Jy) were observed as bandpass and gain calibrators. 
The calibrated visibility data were Fourier-transformed with the Briggs robust parameter of +0.5 and CLEANed with the CASA task ``clean''. 
The synthesized beams of the CLEANed images are 0\farcs92 $\times$ 0\farcs71 for the 1.3 mm continuum, 0\farcs85 $\times$ 0\farcs72 for the $^{12}$CO line, 0\farcs96 $\times$ 0\farcs75 for the C$^{18}$O line, and 0\farcs96 $\times$ 0\farcs74 for the SO line. 
The rms noise levels are $\sim$0.15 mJy beam $^{-1}$ in the 1.3 mm continuum image and $\sim$8 mJy beam$^{-1}$ per channel in the line images. 

\section{Results}
The systemic velocity ($V_{\rm sys}$) of L1489 IRS has been estimated to be 7.2 km s$^{-1}$ in the LSR frame with the single-dish observations in several molecular lines at multiple transitions (Brinch et al.~2007a). 
As will be shown in Section \ref{disk}, 
from our fitting of the Keplerian disk model to the C$^{18}$O velocity channel maps, 
$V_{\rm sys}$ of L1489 IRS is determined to be 7.3 km s$^{-1}$, approximately consistent with $V_{\rm sys}$ measured on thousands of AU scale with the single-dish observations. 
In the present paper, 
$V_{\rm sys}$ = 7.3 km s$^{-1}$ is adopted, 
and all the velocities shown in this paper are the relative velocities ($\tbond V$) with respect to $V_{\rm sys}$.  

\subsection{1.3 mm Continuum Emission}
\begin{figure}
\epsscale{1}
\plotone{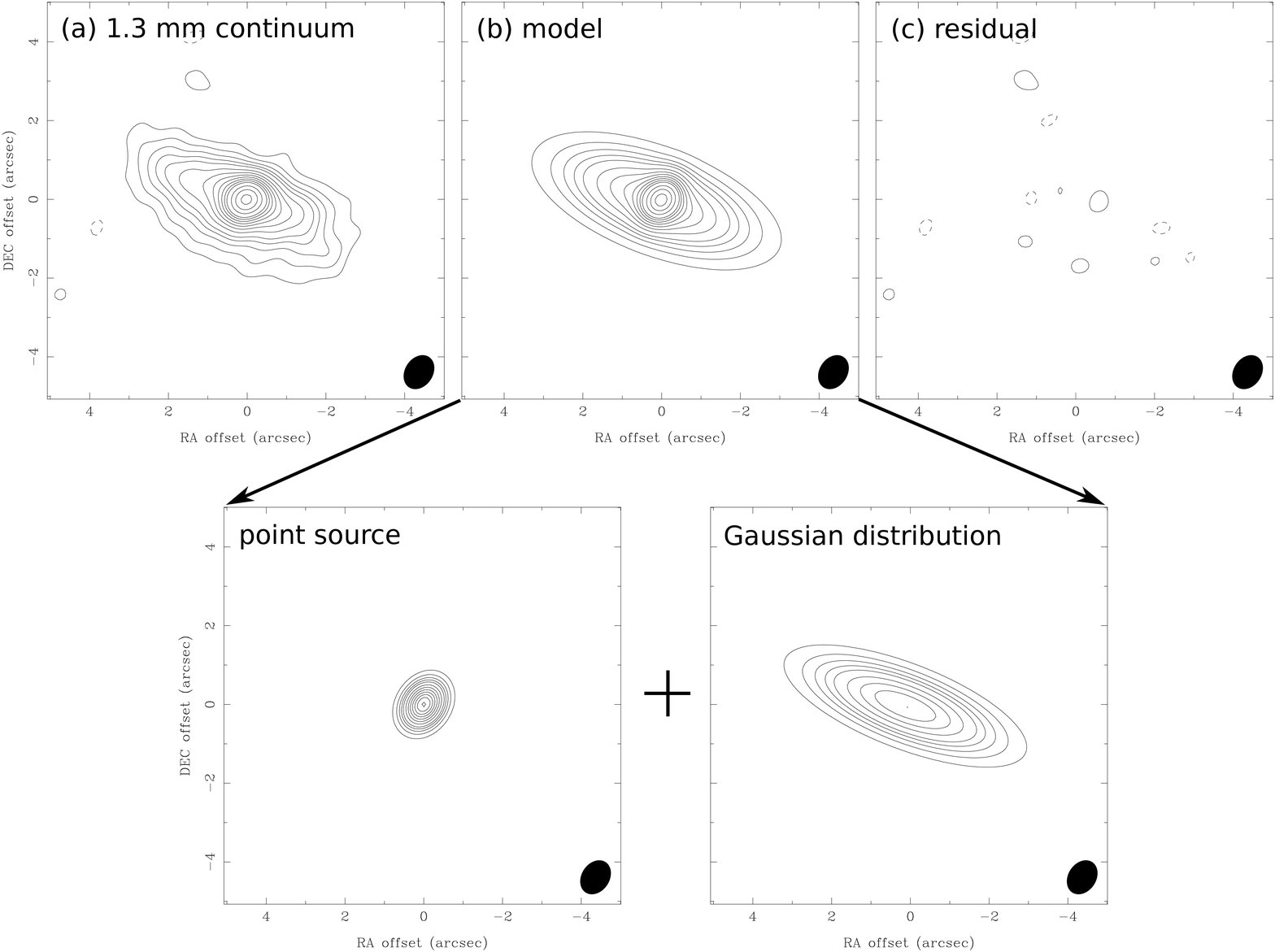}
\caption{(a) 1.3 mm continuum image of L1489 IRS. (b) Model image of the 1.3 mm continuum emission in L1489 IRS, which is composed of a point source and a Gaussian intensity distribution. (c) Residual image after subtracting the model image from the observed continuum image. A filled ellipse at the bottom-right corner in each panel
denotes the beam size. Contour levels are from 3$\sigma$ to 15$\sigma$ in steps of 3$\sigma$, from 15$\sigma$ to 50$\sigma$ in steps of 5$\sigma$, and then from 50$\sigma$ to 80$\sigma$ in steps of 10$\sigma$, where 1$\sigma$ is 0.15 mJy Beam$^{-1}$.}\label{con}
\end{figure}

Figure \ref{con}a shows the observed 1.3 mm continuum image of L1489 IRS. 
The continuum emission is elongated from northeast to southwest, 
and its intensity distribution can be described by a combination of a point source and a Gaussian intensity distribution (Figure \ref{con}b) derived by the MIRIAD fitting program, $imfit$. 
The point source has a total flux of $\sim$7.6$\pm$0.5 mJy and is located at $\alpha$(J2000) = $4^{h}04^{m}43\fs07$, $\delta$(J2000) = $26\arcdeg18\arcmin56\farcs3$. 
In the present paper, this position is adopted as the position of the central protostar. 
The center of the Gaussian component is at $\alpha$(J2000) = $4^{h}04^{m}43\fs09$, $\delta$(J2000) = $26\arcdeg18\arcmin56\farcs3$, consistent with the position of the point source within 0\farcs2. 
The Gaussian component has a total flux of $\sim$42$\pm$3.7 mJy, a deconvolved size of $3\farcs5 \times 1\farcs1$ (500 AU $\times$ 150 AU), and a position angle of $69\degr$. 
The elongation of the Gaussian component is perpendicular to the outflow axis with a position angle of $\sim$165$\degr$ (Hogerheijde et al.~1998). 
Single-dish millimeter continuum observations have shown that the 1.3 mm flux integrated over the 1$\arcmin$ (8400 AU) region around L1489 IRS is $\sim$150 mJy (Motte \& Andr{\'e} 2001), 
and thus one third of the total 1.3 mm continuum flux is originated from the inner 500 AU region traced by the present ALMA observations. 

\subsection{$^{12}$CO (2--1) Emission}
\begin{figure}
\epsscale{1}
\plotone{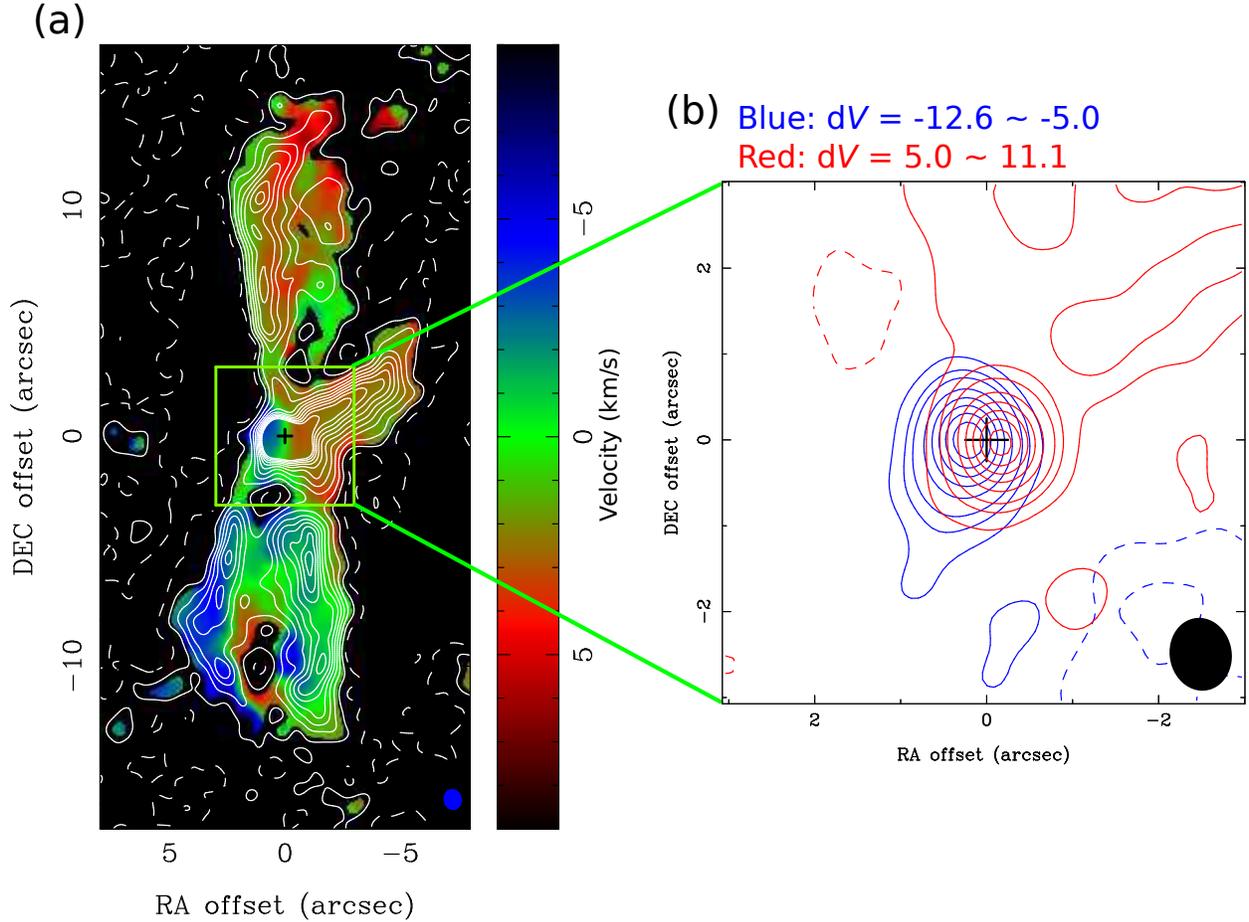}
\caption{(a) Moment 0 map (contour) overlaid on the moment 1 map (color) of the $^{12}$CO (2--1) emission in L1489 IRS. (b) Moment 0 maps of the high-velocity $^{12}$CO (2--1) emission in L1489 IRS. Blue contours show the $^{12}$CO (2--1) emission integrated from $V = -12.5$ km s$^{-1}$ to $V = -5.0$ km s$^{-1}$, and red contours from $V = 5.0$ km s$^{-1}$ to $V = 10.8$ km s$^{-1}$. A green box in (a) shows the area of the high-velocity $^{12}$CO map. A filled ellipse at the bottom-right corner in each panel denotes the beam size. Crosses show the protostellar position. Contour levels are from 5$\sigma$ in steps of 10$\sigma$ in (a), where 1$\sigma$ is 20 mJy Beam$^{-1}$ km s$^{-1}$, and are 5$\sigma$, 15$\sigma$, 25$\sigma$, and then in steps of 20$\sigma$ in (b), where 1$\sigma$ of the integrated blueshifted and redshifted emission is 9 and 8 mJy Beam$^{-1}$ km s$^{-1}$, respectively.}\label{CO}
\end{figure}

Figure \ref{CO}a shows the integrated-intensity (i.e., moment 0) and intensity-weighted mean velocity (i.e., moment 1) maps of the $^{12}$CO (2--1) emission in L1489 IRS. 
The $^{12}$CO emission shows redshifted and blueshifted V-shaped structures with opening angles of $\sim$65$\degr$ and $\sim$35$\degr$ to the north and the south, respectively, 
and their apices are approximately coincident with the protostellar position. 
The elongation and the direction of the velocity gradient of the V-shaped structures are consistent with those of the large-scale outflow as observed with JCMT in the $^{12}$CO (3--2) line (Hogerheijde et al.~1998). 
We consider that the V-shaped structures seen in the $^{12}$CO ALMA observations trace the walls of the outflow cavities.  
In addition to these outflow components, 
there is a central compact component with a size of $\sim$1\farcs7 ($\sim$240 AU) and a positional angle of $\sim$79\degr\ associated with the protostar. 
This central component is elongated perpendicular to the outflow axis and exhibit a velocity gradient along the major axis,   
as clearly seen in the moment 0 maps integrated over the high-velocity ranges ($|V|$ $\gtrsim$ 5 km s$^{-1}$; Figure \ref{CO}b). 
The high-velocity $^{12}$CO emission shows blueshifted and redshifted compact blobs with sizes of $\sim$0\farcs5--0\farcs6 ($\sim$70--80 AU) and positional offsets of $\sim$0\farcs2 ($\sim$30 AU) to the east and the west of the protostar, respectively.
The position angle of the axis passing through the blueshifted and redshifted peak positions of the high-velocity $^{12}$CO emission is $\sim$87$\degr$, perpendicular to the outflow axis. 
The direction of the velocity gradient of the central component is consistent with that of the rotational motion on hundreds of AU scale around the protostar as observed with SMA in the HCO$^+$ (3--2) and C$^{18}$O (2--1) lines (Brinch et al.~2007b; Yen et al.~2013).

\subsection{C$^{18}$O (2--1) Emission}
\begin{figure}
\epsscale{1}
\plotone{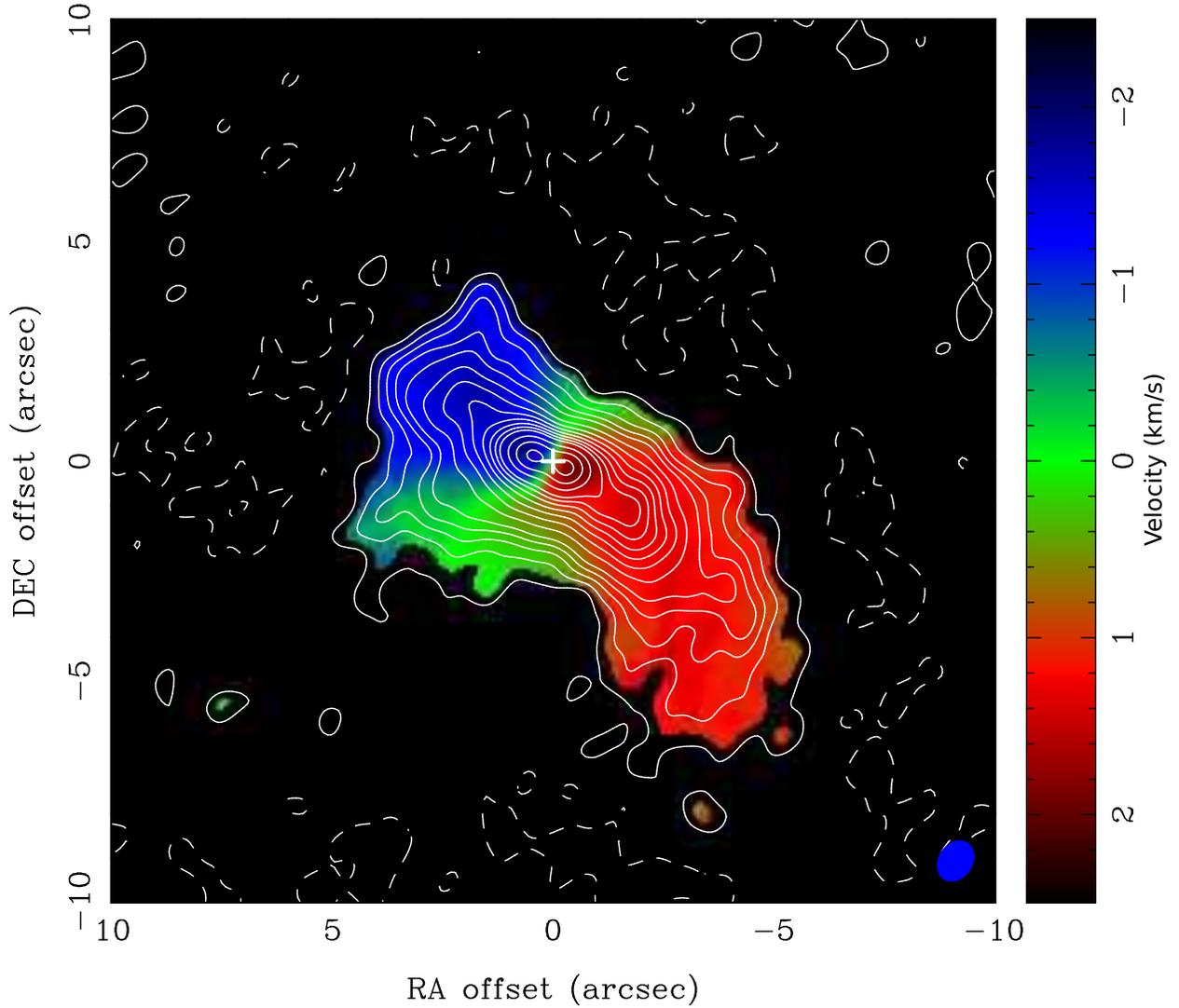}
\caption{Moment 0 map (contour) overlaid on the moment 1 map (color) of the C$^{18}$O (2--1) emission in L1489 IRS. A filled ellipse at the bottom-right corner denotes the beam size. A cross shows the protostellar position. Contour levels are from 3$\sigma$ to 15$\sigma$ in steps of 3$\sigma$, from 15$\sigma$ to 50$\sigma$ in steps of 5$\sigma$, and then from 50$\sigma$ to 90$\sigma$ in steps of 10$\sigma$, where 1$\sigma$ is 10 mJy Beam$^{-1}$ km s$^{-1}$.}\label{C18Omom1}
\end{figure}

Figure \ref{C18Omom1} shows the moment 0 and 1 maps of the C$^{18}$O (2--1) emission in L1489 IRS. 
The C$^{18}$O emission is elongated from northeast to southwest and exhibits a velocity gradient along its major axis (blueshifted to the northeast and redshifted to the southwest), which is consistent with the previous SMA observational results in the HCO$^+$ (3--2) line at an angular resolution of $\sim$1\arcsec\ (Brinch et al.~2007b) and in the C$^{18}$O (2--1) line at an angular resolution of $\sim$4\arcsec\ (Yen et al.~2013). 
In the central region around the protostar, the C$^{18}$O emission shows two peaks, one blueshifted to the east and one redshifted to the west.    
At the protostellar position, the observed mean LSR velocity is close to $V_{\rm sys}$. 
At the tips of the elongated structure, 
there are protrusions of the blueshifted and redshifted emission pointing toward the north and the south, respectively. 
These off-axis protrusions of the blueshifted and redshifted C$^{18}$O emission have also been identified in our previous SMA image of L1489 IRS (Yen et al.~2013). 
By comparing the total C$^{18}$O flux observed with the IRAM 30 m telescope (Hogerheijde et al.~1998),
$\sim$20\% of the total C$^{18}$O flux is recovered with the present ALMA observations. 

\begin{figure}
\epsscale{1}
\plotone{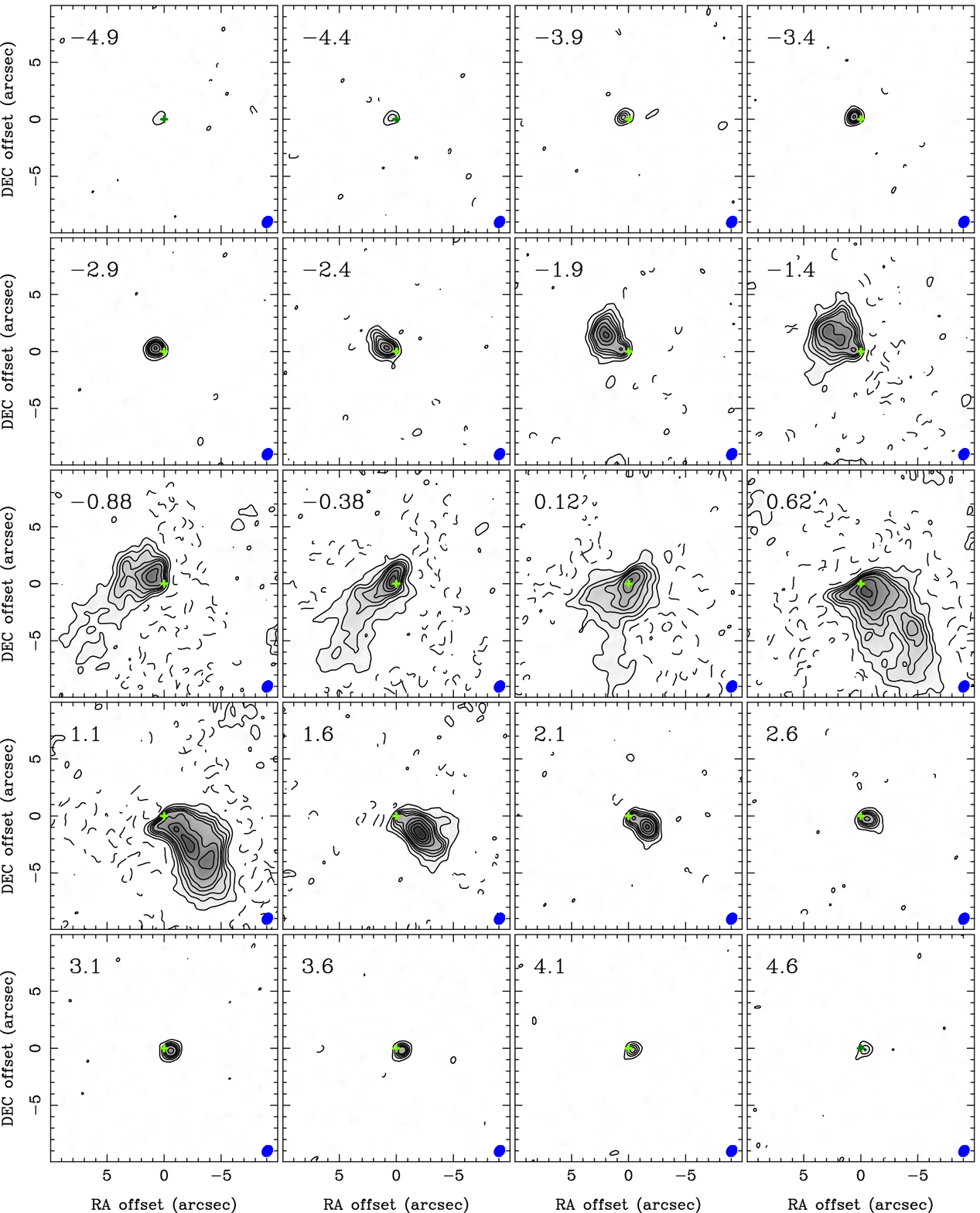}
\caption{Velocity channel maps of the C$^{18}$O (2--1) emission in L1489 IRS (3 channels binned). The central velocity of binned 3 channels is shown at the upper left corner in each panel. A filled ellipse at the bottom-right corner in each panel denotes the beam size. Crosses show the protostellar position. Contour levels are from 3$\sigma$ to 23$\sigma$ in steps of 5$\sigma$, and then in steps of 10$\sigma$, where 1$\sigma$ is 5 mJy Beam$^{-1}$}\label{C18Ochan}
\end{figure}

Figure \ref{C18Ochan} shows the velocity channel maps of the C$^{18}$O emission in L1489 IRS. 
The C$^{18}$O emission is detected above the 3$\sigma$ level over 62 velocity channels from $V$ = $-5.1$ to 5.0 km s$^{-1}$,  
and we binned 3 channels together to reduce the total number of panels for presentation. 
At $V$ $\leq$ $-2.9$ km s$^{-1}$ and $\geq$ 2.6 km s$^{-1}$, the C$^{18}$O emission shows a compact, single-peaked blob to the east and the west of the protostar, respectively.  
From $V$ = $-2.4$ to $-1.4$ km s$^{-1}$, an additional emission component to the northeast of the inner compact blob becomes apparent. 
Around the systemic velocity ($V$ = $-0.88$ to 0.12 km s$^{-1}$), the C$^{18}$O emission is elongated toward the southeast, direction of the blueshifted molecular outflow, 
and thus in these velocities the C$^{18}$O emission is likely contaminated by the outflow. 
The brightest emission peaks at these velocities, however, appear to be associated with the central protostar. 
At $V$ = 0.62 and 1.1 km s$^{-1}$, the C$^{18}$O emission shows an extended, curved structure to the southwest. 
At $V$ =1.6 and 2.1 km s$^{-1}$, two emission components appear; the one closer to the protostar appears to be the same component as that seen in the highest redshifted velocities ($V$ $\geq$ 2.6 km s$^{-1}$), 
and the other is originated from the curved feature.

To present the main features of the C$^{18}$O emission at different velocities, 
in Figure \ref{C18Omap} we show the moment 0 maps of the C$^{18}$O emission integrated over three velocity regimes, high velocities ($V$ = $-5.1$ to $-2.5$ km s$^{-1}$ and 2.5 to 5.0 km s$^{-1}$), medium velocities ($V$ = $-2.4$ to $-1.0$ km s$^{-1}$ and 0.6 to 2.3 km s$^{-1}$), and low velocities ($V$ = $-0.9$ to 0.5 km s$^{-1}$). 
The high-velocity C$^{18}$O emission shows blueshifted and redshifted compact blobs with sizes of $\sim$0\farcs7 ($\sim$100 AU) and positional offsets of $\sim$0\farcs6 ($\sim$80 AU) to the east and the west of the protostar, respectively. 
The position angle of the axis passing through the peak positions of these two blobs is $\sim$70$\degr$, 
consistent with the elongation of the observed 1.3 mm continuum emission. 
The high-velocity $^{12}$CO (Figure \ref{CO}) and C$^{18}$O emission shows similar elongations and directions of the velocity gradients,   
and both likely trace the inner part ($\sim$100 AU scale) of the rotational motion seen on hundreds of AU scale (Brinch et al.~2007b; Yen et al.~2013). 
The medium-velocity C$^{18}$O emission exhibits blueshifted and redshifted peaks located at $\sim$2\arcsec\ ($\sim$280 AU) to the northeast and the southwest, respectively,  
and shows an elongation with a position angle of $\sim$55$\degr$. 
The protrusions seen in Figure \ref{C18Omom1} are clearly seen at the medium velocities.
The low-velocity C$^{18}$O emission shows a central component close to the protostar as well as extensions toward the direction of the blueshifted outflow at both the redshifted and blueshifted velocities. 
The extensions at the low velocities are likely related to the outflow. 
 
\subsection{SO (5$_6$--4$_5$) Emission}
\begin{figure}
\epsscale{1}
\plotone{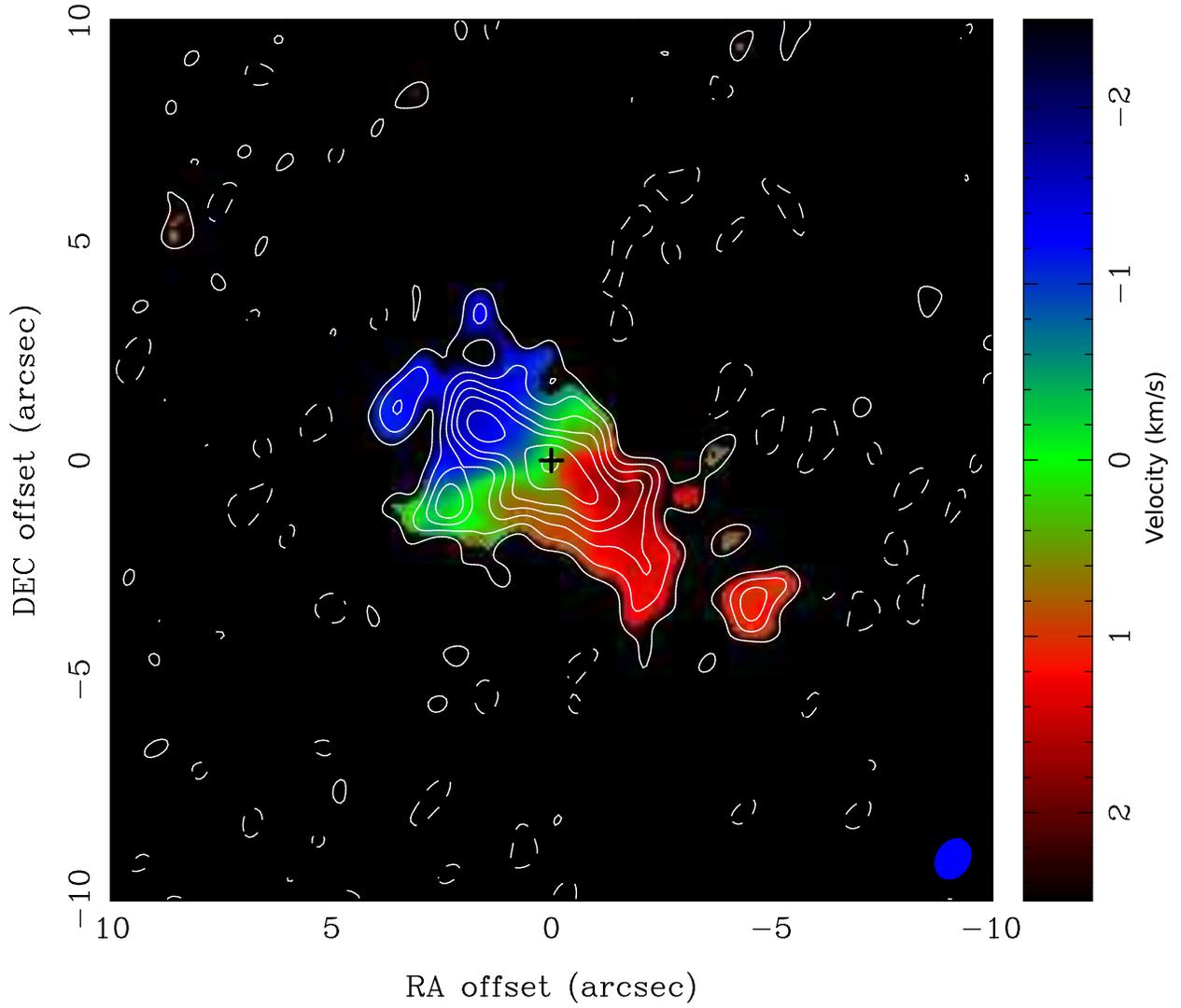}
\caption{Moment 0 map (contour) overlaid on the moment 1 map (color) of the SO (5$_6$--4$_5$) emission in L1489 IRS. A filled ellipse at the bottom-right corner denotes the beam size. A cross shows the protostellar position. Contour levels are from 3$\sigma$ to 15$\sigma$ in steps of 3$\sigma$ and then in steps of 5$\sigma$, where 1$\sigma$ is 8 mJy Beam$^{-1}$ km s$^{-1}$.}\label{SOmom1}
\end{figure}

\begin{figure}
\epsscale{0.8}
\plotone{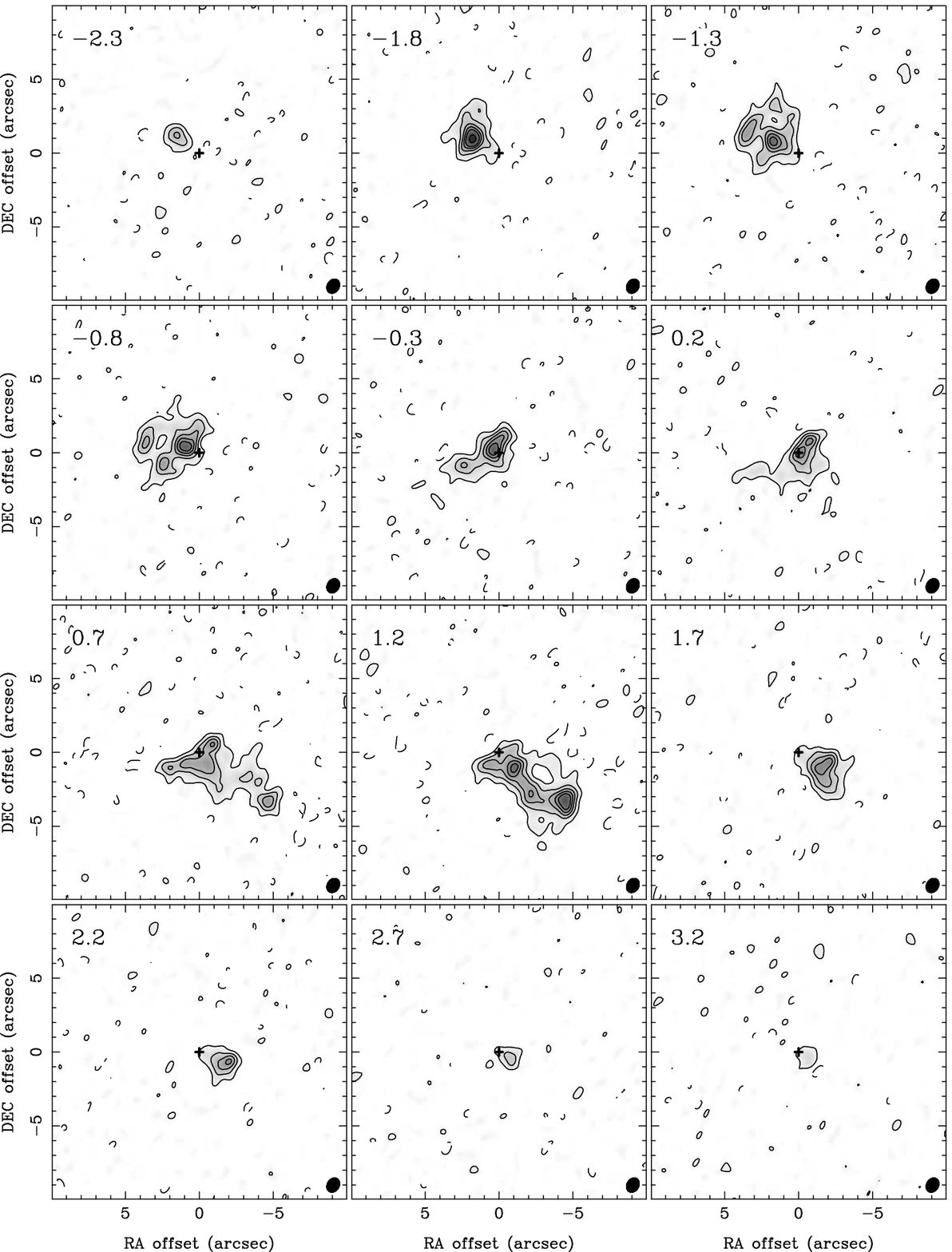}
\caption{Velocity channel maps of the SO (5$_6$--4$_5$) emission in L1489 IRS (3 channels binned). The central velocity of binned 3 channels is shown at the upper left corner in each panel. A filled ellipse at the bottom-right corner in each panel denotes the beam size. Crosses show the protostellar position. Contour levels are from 3$\sigma$ to 18$\sigma$ in steps of 5$\sigma$, and then in steps of 10$\sigma$, where 1$\sigma$ is 5 mJy Beam$^{-1}$.}\label{SOchan}
\end{figure}

Figure \ref{SOmom1} shows the moment 0 and 1 maps of the SO (5$_6$--4$_5$) emission in L1489 IRS. 
The SO emission is elongated from northeast to southwest and exhibits a clear velocity gradient along the major axis as in the case of the C$^{18}$O emission.   
Figure \ref{SOchan} shows the velocity channel maps of the SO emission. 
The SO emission is detected above the 3$\sigma$ levels over 35 velocity channels from $V$ = $-2.6$ to 3.1 km s$^{-1}$, 
and we binned 3 channels together to reduce the total number of the panels for presentation. 
The primary emission peaks at $V$ = $-2.3$ to $-1.3$ km s$^{-1}$ and 1.7 to 2.2 km s$^{-1}$ are coincident with those of the outer components in the C$^{18}$O emission at $V$ = $-1.9$ to $-1.4$ km s$^{-1}$ and 1.6 to 2.1 km s$^{-1}$, 
while the inner C$^{18}$O peaks at a radius of $\sim$0\farcs6 seen in the same velocity range are not identified in the SO emission. 
The SO counterpart of the inner, high-velocity C$^{18}$O component is seen only at the highest redshifted velocities ($V$ = 2.7 to 3.2 km s$^{-1}$). 
At $V$ = $-0.8$ to 0.2 km s$^{-1}$, 
the SO velocity channel maps show a central component plus elongation along the outflow direction, 
similar to the corresponding C$^{18}$O velocity channel maps. 
At $V$ = 0.7 to 1.2 km s$^{-1}$, 
the SO emission is elongated along the northeast--southwest direction, 
and there is an intensity peak at the southwest tip of the elongation. 
This southwest peak is spatially coincident with the arm-like structure seen in the C$^{18}$O emission at $V$ = 0.62 km s$^{-1}$. 

To present the main features of the SO emission and compare these with the C$^{18}$O features, 
in Figure \ref{SOmap} we show the moment 0 maps of the SO emission (contour) integrated over three velocity regimes overlaid on the moment 0 maps of the C$^{18}$O emission (color scale) shown in Figure \ref{C18Omap}. 
At high velocities ($|V|$ $\gtrsim$ 2.5 km s$^{-1}$) no blueshifted SO emission is detected, 
while the redshifted SO emission shows a weak compact blob with a size of $\sim$1\farcs1 ($\sim$150 AU) and a positional offset of $\sim$0\farcs9 ($\sim$130 AU) to the west.
As compared to the high-velocity $^{12}$CO and C$^{18}$O emission, 
the high-velocity SO emission is less compact and located further away from the protostar. 
At medium velocities ($V$ = $-2.5$ to $-1.0$ km s$^{-1}$ and 0.7 to 2.2 km s$^{-1}$), 
the SO emission shows blueshifted and redshifted peaks to the northeast and the southwest, respectively, coincident with the peak positions of the medium-velocity C$^{18}$O emission within 0\farcs1, 
plus an additional redshifted peak further away to the southwest. 
The elongation of the medium-velocity SO emission is consistent with that of the medium-velocity C$^{18}$O emission. 
The redshifted protrusion seen in the medium-velocity C$^{18}$O emission is not clearly identified in the SO emission. 
At the low blueshifted and redshifted velocities ($V$ = $-0.8$ to 0.5 km s$^{-1}$), 
the SO emission also shows central components close to the protostar as well as extensions toward the blueshifted outflow, as in the case of the low-velocity C$^{18}$O emission.   

\section{Analysis}
\subsection{Keplerian Disk around L1489 IRS}\label{disk}
\begin{figure}
\epsscale{1}
\plotone{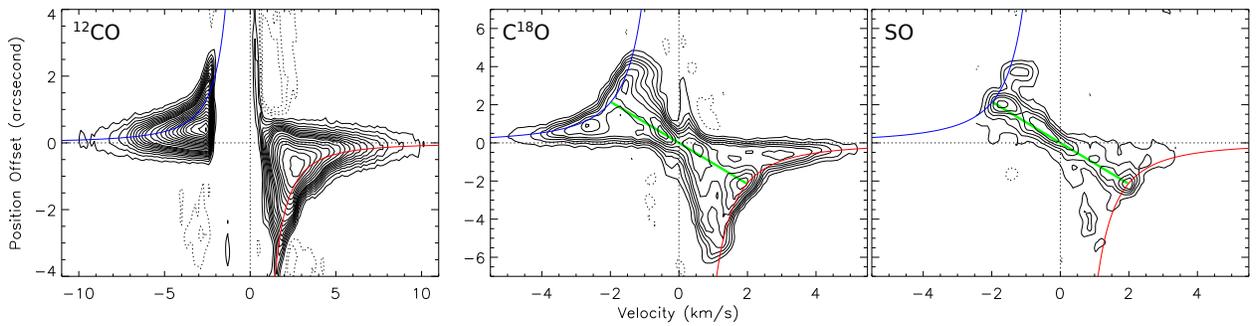}
\caption{P--V diagrams of along the disk major axis (PA = 68\degr) in the $^{12}$CO (left), C$^{18}$O (middle), and SO (right) emission. Blue and red curves show the expected velocities of the Keplerian rotation around a 1.6 $M_\sun$ protostar on the assumption that the inclination angle of the disk plane is 66$\degr$. Green lines delineate the linear velocity gradient seen in the C$^{18}$O and SO emission. Contour levels are from 3$\sigma$ to 15$\sigma$ in steps of 3$\sigma$ and then in 5$\sigma$, where 1$\sigma$ is 8 mJy.}\label{pv}
\end{figure}

The velocity gradients of the $^{12}$CO, C$^{18}$O, and SO emission along the major axis likely trace the rotational motion around L1489 IRS. 
Figure \ref{pv} shows the position--velocity (P--V) diagrams of the $^{12}$CO, C$^{18}$O, and SO emission along the major axis. 
In the P--V diagrams, the velocities of the $^{12}$CO and C$^{18}$O emission increase as the radii decrease, following the same trend as the Keplerian rotation on 100--1000 AU scale observed with SMA in the HCO$^+$ (3--2) and C$^{18}$O (2--1) lines (Brinch et al.~2007b; Yen et al.~2013).   
On the other hand, the SO emission primarily exhibits a linear velocity gradient connecting the two intensity peaks at $V$ $\sim$ $\pm$2 km s$^{-1}$ and positional offsets of $\sim$$\pm$2\arcsec\ (green line in Figure \ref{pv} right), distinct from the  $^{12}$CO and C$^{18}$O emission. 
A corresponding velocity feature can also be identified in the P--V diagram of the C$^{18}$O emission (green line). 

The high-velocity ($|V| \gtrsim$ 2.5 km s$^{-1}$) $^{12}$CO and C$^{18}$O emission most likely traces the inner part ($r \lesssim 200$ AU) of the Keplerian disk. 
The peak brightness temperatures of the $^{12}$CO and C$^{18}$O emission at $|V|$ =  2.5--5 km s$^{-1}$ are $\sim$30--50 K and $\sim$3--5 K, respectively.
The $^{12}$CO brightness temperature is comparable to the typical temperature of disks on a 100 AU scale around T Tauri stars (e.g., P{\'e}rez et al.~2012),
suggesting that the $^{12}$CO emission is likely optically thick and traces the temperature of the inner disk. 
On the assumption of the disk temperature of 30--50 K and the LTE condition, 
the optical depth of the C$^{18}$O emission at velocities of 2.5--5 km s$^{-1}$ is estimated to be $\sim$0.1.    
On the other hand, 
previous infrared and millimeter continuum observations of L1489 IRS have estimated the scale height of the disk at a radius of 100 AU to be 15--25 AU (0\farcs1--0\farcs2; Eisner et al.~2005), 
which is relatively small compared to our spatial resolution ($\sim$130 AU). 
The optically-thin C$^{18}$O emission can probe the entire vertical structures from the mid-plane to the surface of the disk. 
Furthermore, the two-dimensional theoretical calculations of the temperature structures of disks embedded in infalling envelopes with various sound speeds and masses by Visser et al.~(2009) have shown that the temperature at the mid-plane of disks at radii smaller 250 AU is expected to be higher than 20 K, the frozen-out temperature of CO (\"Oberg et al.~2005).
Thus, the high-velocity C$^{18}$O emission arising from the inner 200 AU region is unlikely affected by the frozen-out effect.
Therefore, to derive the physical parameters of the Keplerian disk around L1489 IRS, 
we have built axisymmetric disk models with the geometrically-thin approximation, generated model C$^{18}$O images, and performed $\chi^2$ fitting to the observed channel maps of the high-velocity ($|V| \gtrsim 2.5$ km s$^{-1}$) C$^{18}$O emission.

Our disk model is expressed as, 
\begin{equation}
\Sigma(r) = \Sigma_0 \cdot (\frac{r}{100\ {\rm AU}})^p, 
\end{equation}
\begin{equation}
T(r) = T_0 \cdot (\frac{r}{100\ {\rm AU}})^q, 
\end{equation}
and 
\begin{equation}\label{kep}
V_{\rm rot}(r) = \sqrt{\frac{GM_*}{r}}, 
\end{equation}
where \begin{equation}
r = \sqrt{x^2 + y^2}\cdot d, 
\end{equation}
\begin{equation}
x = \Delta\alpha\cdot\sin{\rm \psi} + \Delta\delta\cdot\cos{\rm \psi}, 
\end{equation}
\begin{equation}
y = (-\Delta\alpha\cdot\cos{\rm \psi} + \Delta\delta\cdot\sin{\rm \psi})/\cos i,  
\end{equation}
$\Sigma(r)$ is the radial profile of the surface number density with the power-low index $p$ projected into the line of sight, 
$T(r)$ is the radial profile of the temperature with the power-law index $q$, 
$V_{\rm rot}(r)$ is the radial profile of the rotational velocities, 
$G$ is the gravitational constant, 
$M_*$ is the central protostellar mass, 
$d$ is the distance of 140 pc, 
$\psi$ is the position angle of the disk major axis, 
$x$ and $y$ are the positional offsets along the disk major and minor axes in the disk plane,  
$\Delta\alpha$ and $\Delta\delta$ are the RA and Dec offsets with respect to the center of mass,  
and $i$ is the inclination angle.
The center of mass ($\alpha_0$ and $\delta_0$) is defined as RA and Dec offsets with respect to the 1.3 mm continuum peak position. 
The line-of-sight velocity ($\tbond$$V_{\rm los}$) is given by, 
\begin{equation}\label{project}
V_{\rm los} = V_{\rm rot} \sin i \cdot \frac{x}{r} + V_{\rm sys}.
\end{equation}
The line profile ($\phi_v$) of the C$^{18}$O line is assumed to be a Gaussian function as, 
\begin{equation}\label{vphi}
\phi_v \propto \exp(-\frac{(v-V_{\rm los})^2}{2{\sigma_v}^2}), 
\end{equation}
where
\begin{equation}
\sigma_v = \sqrt{\frac{2kT(r)}{m} + v_{\rm turb}^2}, 
\end{equation}
$m$ is the C$^{18}$O molecular mass and $k$ is the Boltzmann constant, and $v_{\rm turb}$ is the turbulent dispersion. 
The critical density of the C$^{18}$O emission is $\sim$10$^4$ cm$^{-3}$ and lower than the typical density of protostellar envelopes and disks on hundreds of AU scale ($\gtrsim$10$^5$ cm$^{-3}$), 
and hence the C$^{18}$O emission is likely thermalized (e.g., Yen et al.~2011).
Thus, on the assumption of the LTE condition, 
the C$^{18}$O (2--1) line intensity ($\tbond$$I_\nu$) in our model disk is computed with the radiative transfer equation, 
\begin{equation}\label{rd}
I_{\nu} = B_{\nu}(T) \cdot (1 - \exp^{-\tau_\nu}), 
\end{equation}
where 
\begin{equation}\label{rad}
\tau_\nu = \Sigma(r)X_{\rm C^{18}O}\kappa_\nu\phi_v, 
\end{equation}
$B_{\nu}(T)$ is the Planck function at a temperature of $T$, $X_{\rm C^{18}O}$ is the C$^{18}$O abundance adopted to be 3 $\times$ 10$^{-7}$ (Frerking et al.~1982), and $\tau_\nu$ and $\kappa_\nu$ are the optical depth and absorption coefficient of the C$^{18}$O (2--1) line, respectively. 
The absorption coefficient and population of the C$^{18}$O emission are computed with
\begin{equation}
\kappa_\nu=-\frac{c^2}{8\pi {\nu_0}^2}\frac{g_{J+1}}{g_J}N_J A_{J+1,J} (1-\exp^{-h\nu_0/kT})\phi_v, 
\end{equation}
\begin{equation}\label{Nj}
N_J = \frac{(2J+1)\exp^{-hB_{\rm e} J(J+1)/kT}}{kT/hB_{\rm e}} \cdot N,
\end{equation}
where $h$ is the planck constant, $c$ is the speed of light, $\nu_0$ is the rest frequency, $J$ is the rotational quantum number of the lower energy level, $g_J$ is the statistical weight of the energy level $J$, $N_J$ is the number density at the energy level $J$, $A_{J+1,J}$ is the Einstein coefficient, $B_{\rm e}$ is the rotational constant of 54.89 GHz, and $N$ is the total number density of C$^{18}$O molecules.

The $\chi^2$ fitting is performed on the convolved image cubes. 
We select the central 2$\arcsec$ $\times$ 2$\arcsec$ region and the velocity ranges of $-$6.1 to $-$2.5 and 2.5 to 7.1 km s$^{-1}$, which cover the high-velocity C$^{18}$O emission (Figure \ref{C18Omap} left), as our fitting region.
There are eleven fitting parameters in our disk model ($\alpha_0$, $\delta_0$, $M_*$, $i$, $\psi$, $\Sigma_0$, $p$, $T_0$, $q$, $V_{\rm sys}$, and $v_{\rm turb}$). 
Because we only apply our disk model to the high-velocity C$^{18}$O emission which likely traces the inner part ($r \lesssim 200$ AU) of the Keplerian disk,  
the outer radius of the disk ($\tbond$$R_{\rm d}$) cannot be estimated in this analysis. 
Previous observations of the disk around L1489 IRS have estimated $R_{\rm d}$ to be as large as 200 AU (Brinch et al.~2007b; Eisner 2012). 
Thus, in our disk model, $R_{\rm d}$ is fixed to be 200 AU. 
To reduce the number of free parameters, 
we first adopted the 1.3 mm continuum peak position as the center of mass ($\alpha_0 = \delta_0 = 0$), typical power-law indices of the density and temperature profiles of disks around T Tauri stars ($p = -1.5$ and $q = -0.4$; Guilloteau \& Dutrey 1994; Dutrey et al.~1998; Guilloteau et al.~2011), the elongation of the high-velocity C$^{18}$O emission as the disk major axis ($\psi$ = 70\degr), and the previously reported $V_{\rm sys}$ = 7.2 km s$^{-1}$ (Brinch et al.~2007a). 
For the first run of the fitting,
the fitting parameters are $M_*$, $i$, $\Sigma_0$, $T_0$, and $v_{\rm turb}$, 
and the searching ranges were $1.2 \leq M_* \leq 2.4\ M_\sun$, $30\degr \leq i \leq 80\degr$, $1.5 \times 10^{16} \leq \Sigma_0 \leq 2.4 \times 10^{17}$ cm$^{-2}$, $10 \leq T_0 \leq 60$ K, and $0.2 \leq v_{\rm turb} \leq 0.6$ km s$^{-1}$.
The searching ranges cover the previous reported $M_*$ (1.4 and 1.8 $M_\sun$; Brinch et al.~2007b; Yen et al.~2013) and $i$ (36\degr--74\degr; Kenyon et al.~1993; Padgett et al.~1999; Eisner et al.~2005; Brinch et al.~2007a, b; Eisner 2012) of L1489 IRS, typical density and temperature of disks on the scale of 100 AU around T Tauri stars (e.g., Pi{\'e}tu et al.~2007; Isella et al.~2009; Guilloteau et al.~2011), and subsonic to supersonic turbulent dispersion. 
After the first run of the fitting, 
we obtained initial estimates of $M_*$ = 1.5 $M_\sun$, $i$ = 65$\degr$, $\Sigma_0$ = 6 $\times$ 10$^{16}$ cm$^{-2}$, $T_0$ = 45 K, $v_{\rm turb}$ = 0.3 km s$^{-1}$,  
and the reduced $\chi^2$ of 1.63. 
We then fixed these parameters and adopted $\alpha_0$, $\delta_0$, $\psi$, and $V_{\rm sys}$ as free parameters to repeat the fitting with searching ranges of $-0\farcs3 \leq \alpha_0 \leq 0\farcs3$, $-0\farcs3 \leq \delta_0 \leq 0\farcs3$, $64\degr \leq \psi \leq 76\degr$, and $7.2 \leq V_{\rm sys} \leq 7.5\ {\rm km\ s^{-1}}$. 
The new estimates provided $\alpha_0$ = $0\farcs1$, $\delta_0$ = $-0\farcs1$, $\psi$ = 68$\degr$, $V_{\rm sys}$ = 7.3 km s$^{-1}$, and the reduced $\chi^2$ of 1.22.
Next, we fixed these parameters as well as $p$ and $q$, and adopted $M_*$, $i$, $\Sigma_0$, $T_0$ and $v_{\rm turb}$ as free parameters to repeat the fitting. 
In this fitting run, the best fits of these parameters were $M_*$ = 1.6 $M_\sun$, $i$ = 66$\degr$, $\Sigma_0$ = 6 $\times$ 10$^{16}$ cm$^{-2}$, $T_0$ = 44 K, and $v_{\rm turb}$ = 0.3 km s$^{-1}$, 
and the reduced $\chi^2$ of the fitting is 1.18. 
We have tested the fitting by adopting different initial $p$ from $-1.2$ to $-1.8$ and $q$ from $-0.2$ to $-0.6$, and found that the best-fit results except for $T_0$ are consistent, suggesting that our estimates would not be affected by the initial assumptions of $p$ and $q$. 
Then, we performed fitting of $p$, $q$, and $T_0$ with the other parameters fixed and obtained the best-fit values of $p$ = $-1.7$, $q$ = $-0.3$, and $T_0$ = 44 K with the reduced $\chi^2$ of 1.16 ($\tbond$ $\chi^2_{\rm min}$).

\begin{figure}
\epsscale{0.5}
\plotone{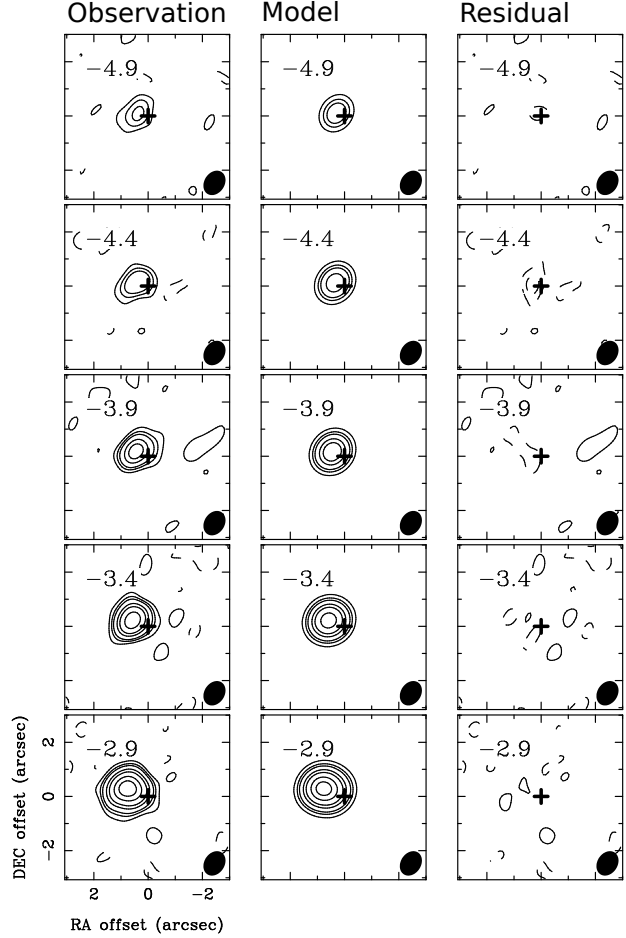}
\caption{Left columns: velocity channel maps of the high-velocity C$^{18}$O (2--1) emission in L1489 IRS. Middle columns: model velocity channel maps of the Keplerian disk traced by the high-velocity C$^{18}$O (2--1) emission. Right columns: residual maps after subtracting the model velocity channel maps from the C$^{18}$O velocity channel maps. We have binned three channels together for presentation. The central velocity of binned 3 channels is at the upper-left corner in each panel. Crosses show the protostellar position. Contour levels are from 2$\sigma$ to 6$\sigma$ in steps of 2$\sigma$ and then in steps of 5$\sigma$, where 1$\sigma$ is 5 mJy Beam$^{-1}$.}\label{diskfit}
\end{figure}

The best-fit results are shown in Figure \ref{diskfit}, and the best-fit parameters are listed in Table \ref{fitting}. 
The best-fit values of $M_*$ and $i$ are consistent with those derived from the previous SMA observations (Brinch et al.~2007b; Yen et al.~2013). 
The levels of of residuals are all $\lesssim$4$\sigma$ and mostly $\lesssim$3$\sigma$,  
except for the velocity channel of $V$ = 3.6 km s$^{-1}$, 
where there are significant residuals of $\sim$6$\sigma$ (Figure \ref{diskfit} right). 
The systematical residual emission component after the subtraction of the best-fit Keplerian-disk model is seen at $V$ = 3.1 to 4.6 km s$^{-1}$. 
The peak position of this emission component does not show systematical change at different velocities, suggesting that this component is unlikely to arise from the Keplerian disk. 
At lower velocities ($V$ = 1.6 to 2.1 km s$^{-1}$; see Figure \ref{C18Ochan}) a second emission component to the southwest of the Keplerian-disk component is clearly seen, 
and thus the residual emission component may be originated from this second component.

The uncertainty of each fitting parameter is estimated by fixing the other parameters at the best-fit values and varying that parameter to explore the profile of the reduced $\chi^2$, 
and we adopt the parameter range that gives a reduced $\chi^2$ of 1+$\chi^2_{\rm min}$ as the fitting uncertainty.
We found that $p$, $q$, $T_0$, and $v_{\rm turb}$ are not well constrained by our fitting. 
As described above, 
with $p$ of $-1.2$ to $-1.8$ and $q$ of $-0.2$ to $-0.6$, 
the best-fit results remain unchanged. 
Even with wider ranges of $p$ (0 to $-3$) and $q$ (0 to $-1$), 
the reduced $\chi^2$ of the fitting is still below 1+$\chi^2_{\rm min}$. 
This is because of the lack of other transitions of C$^{18}$O lines and the insufficient spatial coverage (i.e., only the innermost 200 AU region of the disk observed at the high velocities) to constrain $p$ and $q$ (e.g., Fedele et al.~2013). 
In addition, it is difficult to disentangle intensity changes and line broadening with different $T_0$ because the C$^{18}$O emission is optically thin. 
$v_{\rm turb}$ in our geometrically-thin disk model may not reflect the actual turbulent dispersion, 
but the effect of the scale height of the disk, as will be discussed below. 
In our fitting, we found that when $T_0$ is larger than 29 K and up to 100 K and $v_{\rm turb}$ is within 0--1 km s$^{-1}$, the reduced $\chi^2$ is below 1+$\chi^2_{\rm min}$. 
The uncertainty of each parameter except for $p$, $q$, $T_0$, and $v_{\rm turb}$ is listed in Table \ref{fitting}.

Figure \ref{condisk} presents comparison between the 1.3 mm continuum emission and the high-velocity C$^{18}$O emission. 
The 1.3 mm continuum and the high-velocity C$^{18}$O emission are well aligned and elongated along the same direction. 
Assuming the 1.3 mm continuum emission traces a circular flattened structure, 
the inclination angle of the structure is estimated to be $\sim$72$\degr$ from the length ratio of the minor and major axes, 
which is comparable to the inclination angle of the Keplerian disk derived from the model fitting (= 66\degr).  
Therefore, 
the 1.3 mm continuum emission also likely traces the Keplerian disk. 
The mass traced by the 1.3 mm continuum emission can be estimated as, 
\begin{equation}
M_{\rm d} = \frac{F_{\rm 1.3mm}d^{2}} {\kappa_{\rm 1.3 mm} B(T_{\rm dust})},
\end{equation}
where $F_{\rm 1.3mm}$ is the total 1.3 mm flux, $\kappa_{\rm 1.3 mm}$ is the dust mass opacity at 1.3 mm, and $T_{\rm dust}$ is the dust temperature. 
On the assumption that the wavelength ($\lambda$) dependence of the dust mass opacity ($\equiv \kappa_{\lambda}$) is $\kappa_{\lambda} = 0.1 \times (0.3\ {\rm mm}/\lambda)^{\beta}$ cm$^{2}$ g$^{-1}$ (Beckwith et al.~1990), the mass opacity at 1.3 mm is 0.023 cm$^{2}$ g$^{-1}$ with $\beta = 1.0$ (e.g., J{\o}rgensen et al.~2007). 
The mass of the Keplerian disk is then estimated to be $\sim$0.003--0.007 $M_\sun$ with a dust temperature of 55--25 K, which is derived from our disk model fitting at radii of  50--500 AU.  

\begin{figure}
\epsscale{1}
\plotone{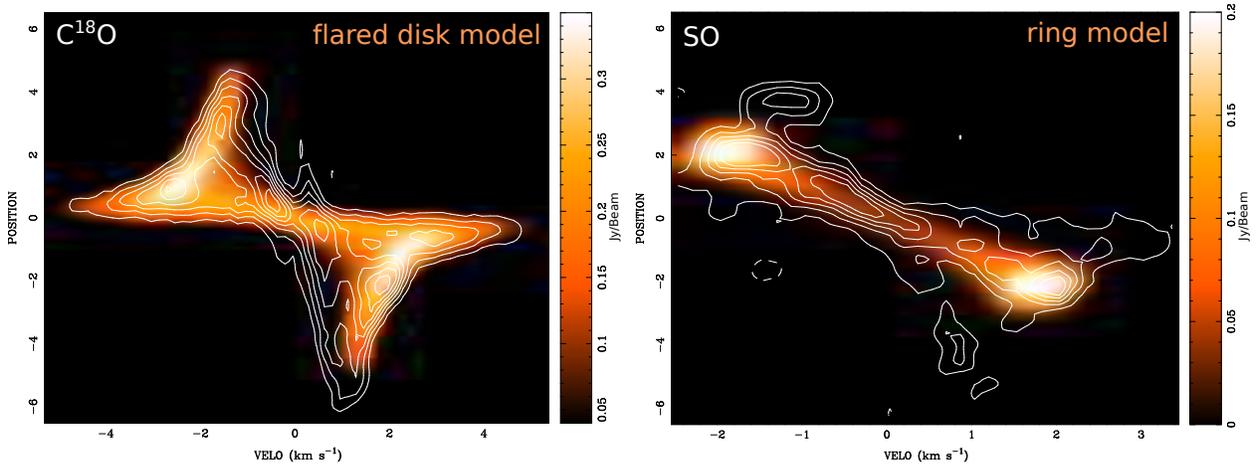}
\caption{P--V diagrams along the disk major axis in the C$^{18}$O (left) and SO (right) emission (contours) overlaid on the model P--V diagrams (color), (a) model of a flared Keplerian disk and (b) model of a ring in the flared Keplerian disk. Contour levels are from 5$\sigma$ in steps of 5$\sigma$ in the C$^{18}$O P--V diagram, and are from 3$\sigma$ in steps of 3$\sigma$ in the SO P--V diagram, where 1$\sigma$ is 8 mJy.}\label{modelpv}
\end{figure}

The P--V diagram of the C$^{18}$O emission (Figure \ref{pv}) shows that along the disk major axis the velocity structure at lower velocities (1.0 $\lesssim |V| \lesssim$ 2.5 km s$^{-1}$) can also be explained by the Keplerian rotation derived by the model fitting of the high-velocity ($|V| \gtrsim$ 2.5 km s$^{-1}$) C$^{18}$O emission (red and blue curves in Figure \ref{pv}). 
This suggests that the radius of the Keplerian disk could be as large as $\sim$700 AU,   
although it is difficult to unambiguously constrain the outer radius of the Keplerian disk due to the effect of the missing flux ($\sim$80\%; Hogerheijde et al.~1998). 
On the other hand, 
the linear velocity gradient seen in the P--V diagram of the C$^{18}$O emission at inner radii ($r \lesssim$ 2$\arcsec$) and lower velocities ($|V| \lesssim$ 2 km s$^{-1}$; green line in Figure \ref{pv}) cannot be explained with our model of a geometrically-thin Keplerian disk. 
Previous infrared and millimeter continuum observations have shown that the disk around L1489 IRS is flared (e.g., Eisner et al.~2005), 
and the disk flaring may account for the linear velocity gradient.  
To compare our disk model and the C$^{18}$O emission at lower velocities ($|V| \lesssim$ 2 km s$^{-1}$), where the scale height is expected to be as large as a few tens of AU, 
we constructed a three-dimensional model of a flared disk based on our best-fit model of the geometrically-thin disk.
We extrapolated our best-fit surface density and temperature profiles to an outer radius of 700 AU ($\sim$5\arcsec, the emission extent seen in the C$^{18}$O P--V diagram), incorporated the scale height, and generated the model images.   
The scale height ($h$) of the model disk is assumed to be 
\begin{equation}\label{scaleh}
h(r) = h_0 \cdot (\frac{r}{100 {\rm AU}})^b, 
\end{equation}
where $h_0$ is adopted to be the previously estimated value, 25 AU (Eisner et al.~2005), and $b$ to be 1.35 from the relation $b = 1.5 + q/2$ (Guilloteau \& Dutrey 1998), which is approximately consistent with the expectation for a flared disk, $b \sim1.29$ (Chiang \& Goldreich 1997).
The density profile is modified from the surface density profile as, 
\begin{equation}\label{rhoprofile}
\rho(r,z) = \rho_0 \cdot (\frac{r}{100 {\rm AU}})^a \exp(-\frac{z^2}{2{h(r)}^2}), 
\end{equation}
\begin{equation}\label{rhoscale}
\rho_0 = \frac{\Sigma_0\cdot \cos i}{\sqrt{2\pi}h_0}, 
\end{equation}
where $a = -3.05$ from the relation $a = -1.5 + p - q/2$ (Guilloteau \& Dutrey 1998), and $\Sigma_0$ and $i$ are 2 $\times$ 10$^{23}$ cm$^{-2}$ and 66$\degr$, respectively, which are the best-fit values derived from our geometrically-thin disk model fitting. 
The C$^{18}$O abundance is adopted to be 3 $\times$ 10$^{-7}$ (Frerking et al.~1982). 
In the case of a flared disk, 
the line of sight tilted with respect to the disk mid-plane intersects different layers of the disk at different radii. 
The radii ($r_l$) and the vertical distances to the mid-plane ($z_l$) of different layers of the disk along the line of sight can be expressed as
\begin{equation}
r_l = l\sin i + r_0,
\end{equation}
\begin{equation}
z_l = l\cos i,
\end{equation}
where $l$ is the distance to the mid-plane along the line of sight and $r_0$ is the disk radius of the intersection between the mid-plane and the line of sight.
The density profile along the line of sight can be evaluated with $r = r_l$, $z = z_l$, and Equation \ref{scaleh}, \ref{rhoprofile}, and \ref{rhoscale}.  
The rotational velocity at $r_l$ and $z_l$ is 
\begin{equation}
V_{\rm rot}(r_l,z_l) = \sqrt{\frac{GM_*r_l}{({r_l}^2+{z_l}^2)}}. 
\end{equation}
For our model of the flared disk,  
we adopt the LTE condition and an isotropic power-law temperature profile with the same power-law index as the best fit of our geometrically-thin disk model, $T(r) = 44 \cdot (r/100\ {\rm AU})^{-0.3}$ K for simplicity.
Such simple model can demonstrate the overall gas distribution and kinematics to compare with our observational data, 
although flared disks are expected to has vertical temperature gradient (e.g., D'Alessio et al.~1998; Dullemond 2002), 
and our model intensities may not be accurate due to the lack of sophisticated temperature structures.
The model images are computed in Cartesian coordinate, 
and the grid size is $\sim$2 AU, 
which is $\sim$60 times smaller than the observational beam size. 
Therefore, the smearing of the velocity structures within a grid is unlikely to affect the prediction of the observed velocity structures. 
The C$^{18}$O emission is detected at velocities $\lesssim$5 km s$^{-1}$, suggesting that it primarily traces the disk at radii $\gtrsim$50 AU, where the scale height is expected to be $\sim$10 AU.
Hence, the grid size in our model is sufficiently small to capture the vertical velocity gradient. 
By comparing our model images of the flared disk and the geometrically-thin disk, 
we found that the C$^{18}$O spectra of the inner part of the flared disk ($r$ $\lesssim$ 200 AU) tend to have a higher peak velocity and a broader line width than those of the geometrically-thin disk, 
which is due to the contribution of the emission from the upper layer of the flared disk at inner radii.  
The broader line width due to the disk flaring can be compensated by the increase in $v_{\rm turb}$ in a geometrically-thin disk model.
The difference in peak velocities of the C$^{18}$O spectra of the flared and geometrically-thin disks is less than 15\% with the range of the inclination angles we searched in our disk fitting (35$\degr$--75$\degr$). 
Therefore, we expect that the protostellar mass can be over estimated in our fitting with a geometrically-thin disk model by $\lesssim$23\%. 
Such uncertainty is comparable to the uncertainty of our disk fitting and does not affect our main discussion and results.
Figure \ref{modelpv} left shows the P--V diagram of our model of the flared disk, 
and the model can reproduce the main velocity structures of the observed C$^{18}$O emission, 
including the linear velocity gradient at the lower velocities ($|V|$ $\lesssim$ 2 km s$^{-1}$).  

\begin{figure}
\epsscale{1}
\plotone{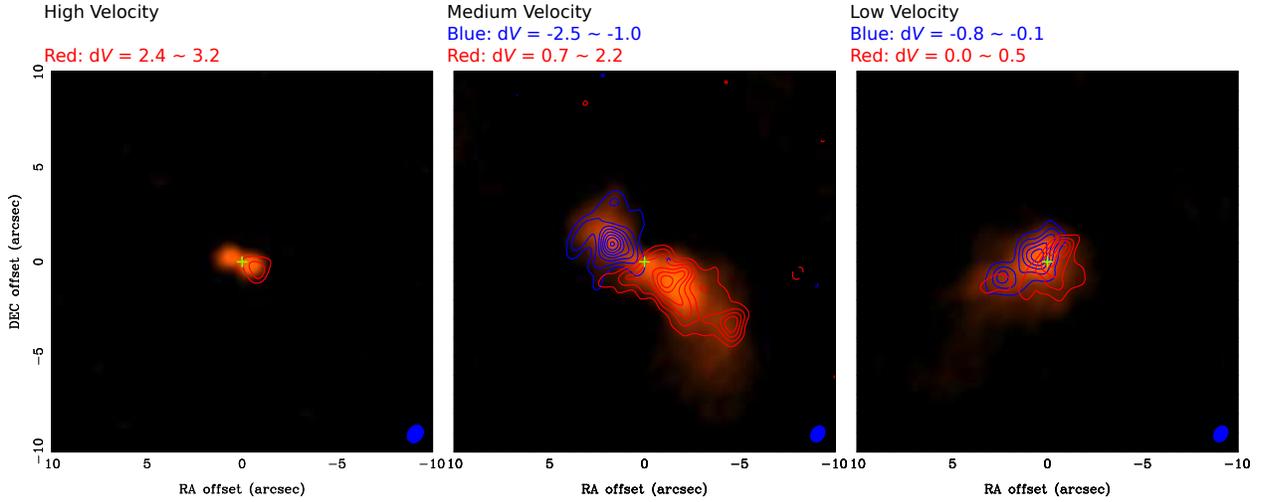}
\caption{Moment 0 map of the high-velocity, medium-velocity, and low-velocity SO (5$_6$--4$_5$) emission (contour) in L1489 IRS overlaid on the moment 0 maps of the C$^{18}$O emission integrated over almost the same velocity ranges (color scale). The integrated velocity ranges are shown above the relevant panels. A filled ellipse at the bottom-right corner in each panel denotes the beam size. Crosses show the protostellar position. Contour levels are from 5$\sigma$ in steps of 5$\sigma$, where 1$\sigma$ is 3, 5, and 3 mJy Beam$^{-1}$ km s$^{-1}$ in the high-velocity, medium-velocity, and low-velocity maps, respectively.}\label{SOmap}
\end{figure}

The P--V diagram of the SO emission predominantly shows the linear velocity gradient without clear high-velocity compact emission (Figure \ref{pv}), 
suggesting that the SO emission does not trace the inner part of the Keplerian disk.  
The moment 0 maps of the C$^{18}$O and SO emission at lower velocities ($|V| \lesssim$ 2.5 km s$^{-1}$) show the similar elongation and peak positions (Figure \ref{SOmap}),  
suggesting that the SO emission traces similar structures to those of the C$^{18}$O emission at the lower velocities. 
One possibility of the nature of the SO emission is that it primarily arises from a ring-like region in the flared Keplerian disk around L1489 IRS. 
Equation \ref{project} shows that $V_{\rm los} \propto x$ if $r$ and $V_{\rm rot}$ are fixed. 
Therefore, 
a ring in a Keplerian disk should show a linear velocity gradient in its P--V diagram along the disk major axis.  
Furthermore, 
the two primary SO intensity peaks at positional offsets $\sim$$\pm$2$\arcsec$ in the P--V diagram can be interpreted as the rim-brightened edges of the ring.
To demonstrate this interpretation, we adopt the same flared disk model described above, 
and assume that the SO emission traces only a certain ring region in the flared disk. 
Assuming that the two SO emission peaks at positional offsets of $\sim$$\pm$2$\arcsec$ and velocities of $\sim$$\pm$2 km s$^{-1}$ trace the rims of the ring on both sides, 
the inner and outer radii of the SO ring are roughly estimated to be $r$ $\sim$ 250 and 390 AU. 
Our flared disk model shows that the gas densities and temperatures in this ring region range from 10$^6$ to 10$^7$ cm$^{-3}$ and 20 to 30 K, respectively. 
The peak brightness temperatures of the SO emission are $\sim$3--7 K. 
With the statistical equilibrium radiative transfer code, RADEX (Van der Tak et al. 2007), the optical depths of the SO emission at these densities and temperatures are estimated to be $\sim$0.2--0.6. 
Thus, in the ring region both the C$^{18}$O and SO emission are optically thin, and we simply scaled the model C$^{18}$O image cubes by the SO/C$^{18}$O intensity ratio ($\sim$0.6) in the radial range of the ring, and outside of the radial range we assume no SO emission.
Such simple model can demonstrate the gas distribution and kinematics of the ring model to compare with observational data.
In Figure \ref{modelpv}, the right panel presents the P--V diagrams of the observed SO emission and our model of the ring in the flared Keplerian disk. 
The linear velocity gradient seen in the SO emission as well as the intensity peaks at positional offsets of $\sim$$\pm$2$\arcsec$ and velocities of $\sim$$\pm$2 km s$^{-1}$ can be reproduced with our ring model. 
These results show that the SO emission primarily traces the ring structure in the flared disk. 
We note that in the observed SO P--V diagram there is another intensity peak at a positional offset of $\sim$1$\arcsec$ and a velocity of $\sim$$-0.5$ km s$^{-1}$ without any corresponding intensity peak in our axisymmetric model P--V diagram. 
This additional SO emission peak is likely due to the contamination from the associated molecular outflow, since the SO velocity channel maps around $V_{\rm sys}$ (Figure \ref{SOchan} and \ref{SOmap} right) show that the SO emission is elongated along the outflow direction from the protostellar position.

\subsection{Non-Keplerian Components: Infalling Flows}\label{flow}
As shown in the last subsection, 
the high-velocity C$^{18}$O emission most likely traces the inner part of the Keplerian disk,   
and the P--V diagram of the C$^{18}$O emission along the major axis is reproduced by our model of the flared Keplerian disk. 
On the other hand, 
the moment 0 map of the medium-velocity C$^{18}$O emission shows that there are blueshifted and redshifted protrusions offset from the major axis and pointing toward the north and the south, respectively. 
These ``off-axis'' non-axisymmetric components cannot be explained by any simple axisymmetric disk model. 
Furthermore, 
these protrusions are unlikely to be related with the outflow, 
because the direction of the velocity gradient of the protrusions (blueshifted to the north and redshifted to the south) is opposite to that of the outflow (redshifted to the north and blueshifted to the south).

We here examine the spectra along the ridges of these protrusions to understand their kinematics. 
To extract the spectra along the ridges, we first made the transverse intensity profiles of the protrusions by cutting along the RA or Dec directions at an half beam interval, as shown by black dashed lines in Figure \ref{infall}. 
Then we measured the peak positions of the intensity profiles (green filled squares in Figure \ref{infall}), and extracted the C$^{18}$O spectra at those peak positions. 
Figure \ref{spec} shows the two and four representative C$^{18}$O spectra in the blueshifted and redshifted protrusions, respectively. 
Assuming that the protrusions and the Keplerian disk are coplanar, the expected line-of-sight velocities originated from the Keplerian rotation at these peak positions can be calculated with Equations 3--7, which are shown with vertical dotted lines in Figure \ref{spec}. 
At point A and B in the blueshifted protrusion, the spectra show that the peak velocities are $-1.9$ and $-1.6$ km s$^{-1}$, whereas the expected Keplerian velocities are $-1.3$ and $-0.5$ km s$^{-1}$, respectively. 
Therefore, 
the velocities of the blueshifted protrusions are 0.5--1.0 km s$^{-1}$ more blueshifted with respect to the expected Keplerian velocities. 
Similarly, the peak velocities of the spectra in the redshifted protrusions are 1.7, 1.3, 1.0, and 1.0 km s$^{-1}$ at point C, D, E, and F, 0.2--0.6 km s$^{-1}$  more redshifted than the expected Keplerian velocities of 1.5, 0.9, 0.6, and 0.4 km s$^{-1}$, 
respectively. Furthermore, the deviations of the observed peak velocities from the expected Keplerian velocities become larger as the positions are further away from the protostar (i.e., from point C to F, the velocity deviation increases from 0.2 to 0.6 km s$^{-1}$). 
The effect of the optical depth is unlikely to explain these systematic deviations, because from the C$^{18}$O (2--1)/(3--2) line ratio of $\sim$1 observed with the IRAM 30m telescope at angular resolution of 8$\arcsec$--11$\arcsec$ ($\sim$1100--1500 AU; Hogerheijde et al.~1998), comparable to the size scales of the protrusions, 
the optical depth of the C$^{18}$O (2--1) line is estimated to be $\sim$0.3 at the spectral peaks. 
These results imply that these protrusions are distinct kinematical components from the Keplerian disk.

\begin{figure}
\epsscale{1}
\plotone{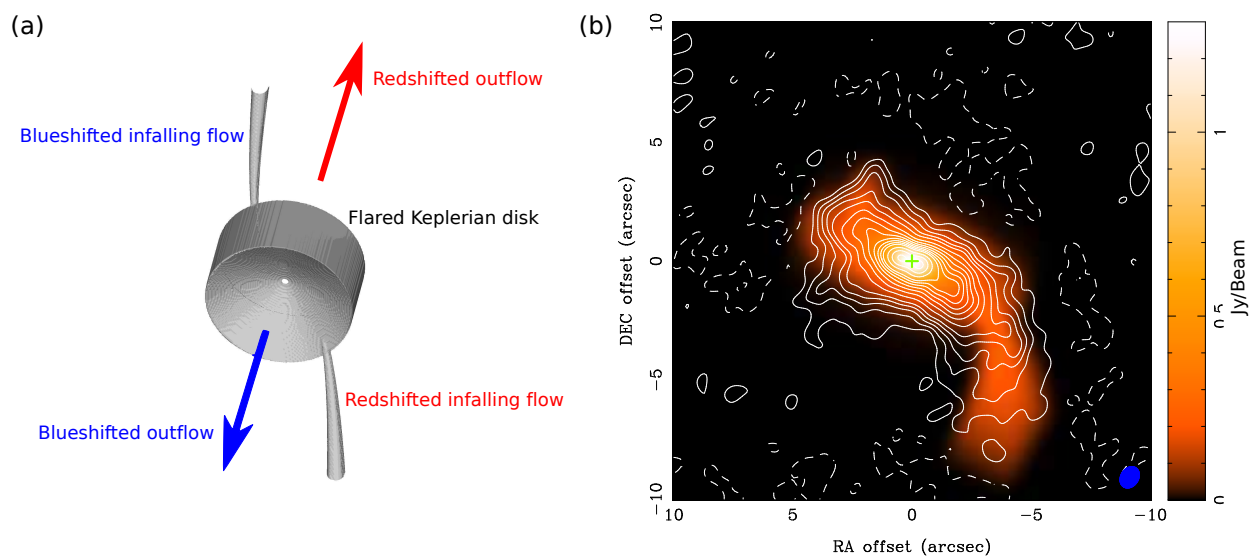}
\caption{(a) Schematic figure demonstrating the configuration of our model of a flared Keplerian disk with two streams of infalling flows toward the disk. Blue and red arrows show the directions of the blueshifted and redshifted outflows, respectively. (b) Moment 0 map of the C$^{18}$O emission (contour; same as Figure \ref{C18Omom1}) overlaid on the model moment 0 map of the flared Keplerian disk with the infalling flows (color scale). A filled ellipse at the bottom-right corner denotes the beam size. A cross shows the protostellar position.}\label{config}
\end{figure}

One possibility of the nature of the non-Keplerian protrusions is that they trace infalling motion toward the Keplerian disk. 
When envelope material falls with a conserved angular momentum and zero total energy, 
it should follow a parabolic trajectory, 
which could form bent structures as seen in the observed protrusions. 
We have calculated such parabolic trajectories of material free-falling from different incoming polar angles toward a 1.6 $M_\sun$ protostar, 
and searched for the best-fit trajectories that trace the measured peak positions (green filled squares in Figure \ref{infall}) and reproduce the curvatures of the protrusions. 
On the assumption that the gravity is dominated by the central protostar,  
parabolic trajectories of infalling material are described as, 
\begin{equation}\label{rin}
r = \frac{R_{\rm c}\cos\theta_0\sin^2\theta_0}{\cos\theta_0-\cos\theta}
\end{equation}
where 
\begin{equation}
\cos\theta = \cos\theta_0\cos\alpha, 
\end{equation}
\begin{equation}
\tan(\phi-\phi_0) = \tan\alpha\sin\theta_0.  
\end{equation}
In the above expression $R_{\rm c}$ is the centrifugal radius, $\theta_0$ and $\phi_0$ are the initial polar and azimuthal angles of the incoming infalling material, and $\alpha$ is the angle between the initial and current position of the infalling material measured at the protostellar position (Ulrich 1976). 
Theoretical models of an infalling envelope have shown density enhancement at the centrifugal radius, where the infalling material falls onto the Keplerian disk (Ulrich 1976; Cassen \& Moosman 1981; Terebey et al.~1984). 
In the moment 0 maps of the medium-velocity C$^{18}$O emission (Figure \ref{C18Omap}), 
there are blueshifted and redshifted intensity peaks adjunct to the protrusions. 
Similar intensity peaks are also seen in the medium-velocity SO emission (Figure \ref{SOmap}). 
If the non-Keplerian protrusions trace the infalling motion toward the Keplerian disk, 
the intensity peaks adjunct to the protrusions could be related to the density enhancement at the centrifugal radius. 
Hence, in our calculation, we adopted the deprojected radius of these intensity peaks ($\sim$300 AU on the assumption the inclination angle of the disk plane $\sim$66\degr) as the centrifugal radius of the infalling material, i.e., $R_{\rm c}$ = 300 AU, and $\theta_0$ and $\phi_0$ as free parameters.
Our calculation shows that the curvature of the blueshifted protrusion can be explained by a parabolic trajectory of material infalling from $\theta_0$ = 52$\degr$ and $\phi_0$ = 74$\degr$ (blue curve in Figure \ref{infall} left), and that of the redshifted protrusion $\theta_0$ = 124$\degr$ and $\phi_0$ = 262$\degr$ (red curve in Figure \ref{infall} right). 
The derived infalling trajectories and the Keplerian disk are not coplanar, 
and the blueshifted and redshifted infalling trajectories are in the far and near sides of the disk plane, respectively (see Figure \ref{config}a). 

Next, we examine whether the observed velocity structures of the protrusions match with the predicted motions of the free-falling gas. 
We have constructed models of two streams of infalling flows along the parabolic trajectories derived above, combined our models of the infalling flows and the flared Keplerian disk, and generated model images.   
The velocities and density distributions of the infalling flows are described as (e.g., Ulrich 1976; Mendoza et al.~2004), 
\begin{equation}\label{vin}
V_r = -\sqrt{\frac{GM_*}{r}}\cdot\sqrt{1+\frac{\cos\theta}{\cos\theta_0}}, 
\end{equation}
\begin{equation} 
V_\theta = \sqrt{\frac{GM_*}{r}}\cdot\frac{(\cos\theta_0-\cos\theta)}{\sin\theta}\cdot\sqrt{1+\frac{\cos\theta}{\cos\theta_0}}, 
\end{equation}
\begin{equation} 
V_\phi = \sqrt{\frac{GM_*}{r}}\cdot\frac{\sin\theta_0}{\sin\theta}\cdot\sqrt{1-\frac{\cos\theta}{\cos\theta_0}}, 
\end{equation}
\begin{equation}
n = n_1 \cdot (\frac{r}{R_{\rm c}})^{-1.5} (1 + \frac{\cos\theta}{\cos\theta_0})^{-0.5} (\frac{\cos\theta}{2\cos\theta_0} + \frac{R_{\rm c}}{r}\cos^2\theta_0)^{-1},    
\end{equation}
where $R_{\rm c}$ is adopted to be 300 AU, $\theta_0$ = 124$\degr$ and $\phi_0$ = 262$\degr$ for the redshifted infalling trajectory, and $\theta_0$ = 52$\degr$ and $\phi_0$ = 74$\degr$ for the blueshifted infalling trajectory. 
In reality the infalling flows must have certain characteristic widths, which depend on the distribution of the injected material as a function of $\theta_0$ and $\phi_0$,  
but it is not straightforward to unambiguously determine such widths and density distribution. 
For simplicity, in our model we assume that both redshifted and blueshifted infall flows have an initial solid angle of $\theta_0\pm3\degr$ and $\phi_0\pm3\degr$, $n_1$ of 3 $\times$ 10$^{7}$ cm$^{-3}$, and C$^{18}$O abundance of 3 $\times$ 10$^{-7}$, which produce the model image approximately consistent with the observed images. 
The lengths of the redshifted and blueshifted infalling flows are adopted to be 5000 AU and 2000 AU, respectively. 
The temperature profile is assumed to be isotropic and extrapolated from the best-fit temperature profile of the Keplerian disk, $T(r) = 44 \cdot (r/100\ {\rm AU})^{-0.3}$ K.
Figure \ref{config} compares the observed and model moment 0 maps, 
and Figure \ref{modelmap}a and b present the P--V diagrams of the observed C$^{18}$O emission and the model along the curvature of the protrusions (blue and red curves in Figure \ref{infall}). 
For comparison, 
we have also made a model of Keplerian-rotating protrusions which are coplanar to the central Keplerian disk (Figure \ref{modelmap}c and d). 
Figure \ref{modelmap} shows that the velocity structures of the protrusions can be explained by the infalling model better than the Keplerian rotation model. 

Our model of infalling flows suggests presence of infalling gas which is not co-planar to the inner Keplerian disk, 
and the angles between the infalling flows and the disk plane range 30$\degr$--40$\degr$. 
Previous SMA observations of L1489 IRS in the millimeter continuum and HCO$^+$ (3--2) line by Brinch et al.~(2007b) and their modeling also show the infalling envelope tilted by 34$\degr$ from the central disk plane, consistent with our results. 
The nature of the infalling flows will be further discussed in the next section.

\section{Discussion}
Our ALMA results and analyses show that L1489 IRS is surrounded by a Keplerian disk with a radius of 700 AU plus non-Keplerian protrusions seen in the C$^{18}$O emission on a 1000 AU scale.  
These protrusions likely trace two streams of infalling flows from the north and the south, which are adjunct to the eastern and western parts of the Keplerian disk, respectively. 
Their velocity structures can be modeled by parabolic free-fall motion with a conserved angular momentum.
These results suggest that the central Keplerian disk is fed material through the non-axisymmetric gas flows.
On larger scales from thousands of AU to 0.1 pc, non-axisymmetric structures of protostellar envelopes, such as filamentary structures, have been proposed from 8 $\mu$m extinction maps (Tobin et al.~2010) and millimeter observations of protostellar envelopes (Tobin et al.~2011). 
For such non-axisymmetric envelope structures it is not straightforward to disentangle between the gas structures and kinematics (Tobin et al.~2012b). 
On the other hand, on smaller scales of $\sim$1000 AU, 
the envelope structures appear to be more or less symmetric (Tobin et al. 2011; 2012a,b),
and previous observations of Class 0 and I sources without sizable Keplerian disks with radii of hundreds of AU, such as B335 (Yen et al.~2010), HH 212 (Lee et al.~2006), L1527 IRS (Ohashi et al.~1997), L1551 IRS 5 (Momose et al.~1998), and IRAS 16293$-$2422 (Takakuwa et al.~2007), show that the kinematics of their protostellar envelopes on a 1000 AU scale can be explained by axisymmetric models of infalling and rotational motions. 
On the other hand, 
our ALMA observations show that in L1489 IRS, which is more evolved than the sources mentioned above, on a 1000 AU scale the infalling material is primarily from the two directions (north and south) and cannot be reproduced by an axisymmetric model.  
The presence of the non-axisymmetric infalling motion in L1489 IRS could be due to its later evolutionary stage and the density distribution of the surrounding envelope. 
Previous observations have found that protostellar envelopes around more evolved sources tend to be less massive (e.g., Ladd et al.~1998;  Arce \& Sargent 2006, J{\o}rgensen et al.~2009), and that opening angles of outflow cavities increase as protostellar sources evolve to later stages (Velusamy \& Langer 1998; Arce \& Sargent 2004, 2006).  
These observational results suggest that protostellar envelopes can be distorted and cleared away by outflows in later evolutionary stages. 
Single-dish observations in N$_2$H$^+$ (1--0) and H$^{13}$CO$^+$ (1--0) lines at angular resolutions of 54$\arcsec$ and 20$\arcsec$, respectively, found no large-scale ($\sim$0.1 pc) dense core associated with L1489 IRS (Caselli et al.~2002; Onishi et al.~2002). 
Observations in the HCO$^+$ (4--3, 3--2, and 1--0) lines with JCMT and the IRAM 30 m telescope at an angular resolution of 14\arcsec--28$\arcsec$ show that L1489 IRS is surrounded by an envelope with a radius of $\sim$30$\arcsec$ (4200 AU; Hogerheijde et al.~1997). 
This envelope is resolved with OVRO observations at an angular resolution of $\sim$5$\arcsec$ (Hogerheijde 2001), 
and most of the envelope material is distributed in the north and the south, 
the directions where the infalling flows come from. 
These observational results suggest that most of the ambient gas around L1489 IRS has been dissipated,  
and that the density distribution of the remaining envelope around L1489 IRS and hence the infalling motion are asymmetric and not co-planar to the central disk. 

Here we discuss whether the identified infalling flows can further grow the central Keplerian disk or not. 
The key physical parameters are specific angular momenta of the infalling material and the central disk, mass infalling rate, and the remaining envelope mass. 
The specific angular momentum of the Keplerian disk can be evaluated as $\sqrt{GM_*R_{\rm d}}$, where $R_{\rm d}$ denotes the outer radius of the Keplerian disk. 
The P--V diagram of the C$^{18}$O emission along the major axis (Figure \ref{pv}) shows that the Keplerian rotation profile extends to a radius of $\sim$5$\arcsec$ ($\sim$700 AU), 
and with $R_{\rm d}$ of 700 AU and $M_*$ of 1.6 $M_\sun$ the specific angular momentum of the Keplerian disk is estimated to be 4.8 $\times$ 10$^{-3}$ km s$^{-1}$ pc$^{-1}$. On the other hand, 
the specific angular momentum of the infalling flows is estimated as $\sqrt{GMR_{\rm c}}/\sin\theta_0$ $\sim$ 2.6 $\times$ 10$^{-3}$ km s$^{-1}$ pc$^{-1}$ with $R_{\rm c}$ of 300 AU and $\theta_0$ of 52$\degr$ or 124$\degr$. 
Considering the uncertainties of $R_{\rm d}$, $R_{\rm c}$, and $M_*$ due to the effect of the missing flux and the limited spatial resolution, 
the specific angular momenta of the Keplerian disk and the infalling flows are comparable within the attainable observational accuracy.
In the theoretical models of collapsing cores without magnetic field (Ulrich 1976; Cassen \& Moosman 1981; Terebey et al.~1984; Basu 1998), 
material infalling from an outer envelope has a angular momentum higher than that in the outer part of the disk, 
and hence the disk gains angular momenta from the infalling material and expands its size as the collapse proceeds. 
In L1489 IRS, 
it is possible that the infalling material can supply sufficient angular momenta to the Keplerian disk to enable the disk growth, 
if the specific angular momenta of the infalling flows is indeed higher.
On the other hand, 
it is also possible that the infalling gas has specific angular momenta lower than those at the outer radius of the Keplerian disk and falls onto the intermediate part of the disk rather than the outer boundary. 
Indeed, the identified infalling flows and Keplerian disk are not co-planar, 
and the landing point of the infalling flows ($\sim$300 AU) can be smaller than the inferrer outer radius of the Keplerian disk.
In this case, the disk can expand through the redistribution of the angular momentum within the disk as the disk material is accreted onto the central protostar.

\begin{figure}
\epsscale{1}
\plotone{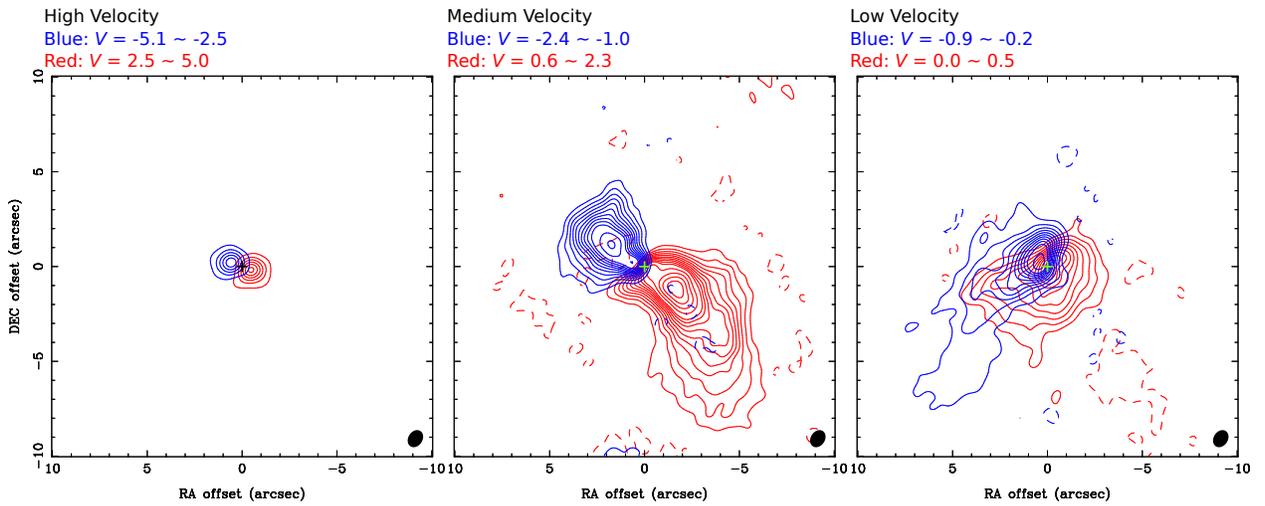}
\caption{Moment 0 map of the high-velocity, medium-velocity, and low-velocity C$^{18}$O (2--1) emission in L1489 IRS. The integrated velocity ranges are shown above the relevant panels. A filled ellipse at the bottom-right corner in each panel denotes the beam size. Crosses show the protostellar position. Contour levels are from 5$\sigma$ in steps of 10$\sigma$ in the high-velocity map, and are from 5$\sigma$ to 50$\sigma$ in steps of 5$\sigma$ and then in steps of 10$\sigma$ in the medium- and low-velocity maps. 1$\sigma$ is 5, 6, 4, 5, 3 and 2 mJy Beam$^{-1}$ km s$^{-1}$ in the high-velocity blueshifted, high-velocity redshifted, medium-velocity blueshifted, medium-velocity redshifted, low-velocity blueshifted, and low-velocity redshifted maps, respectively.}\label{C18Omap}
\end{figure}

\begin{figure}
\epsscale{0.5}
\plotone{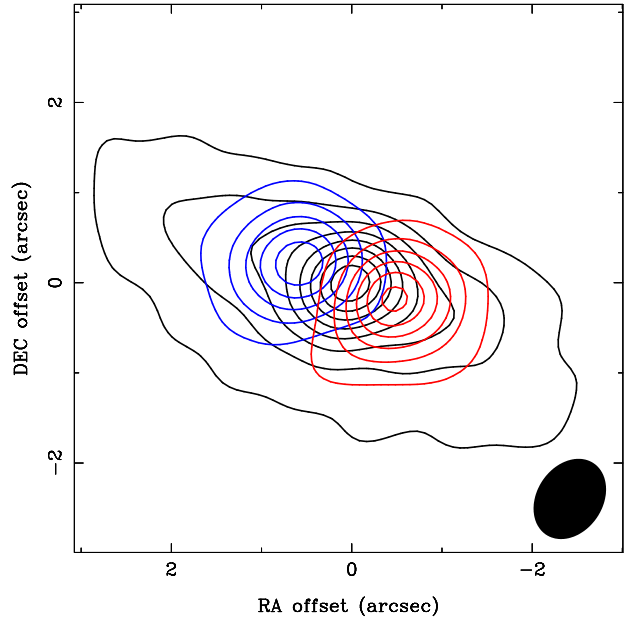}
\caption{Comparison between the 1.3 mm continuum emission (black) with the high-velocity blueshifted and redshifted C$^{18}$O emission, (same as Figure \ref{C18Omap} left). A filled ellipse shows the beam size. Contour levels are from 5$\sigma$ in steps of 5$\sigma$.}\label{condisk}
\end{figure}

\begin{figure}
\epsscale{1}
\plotone{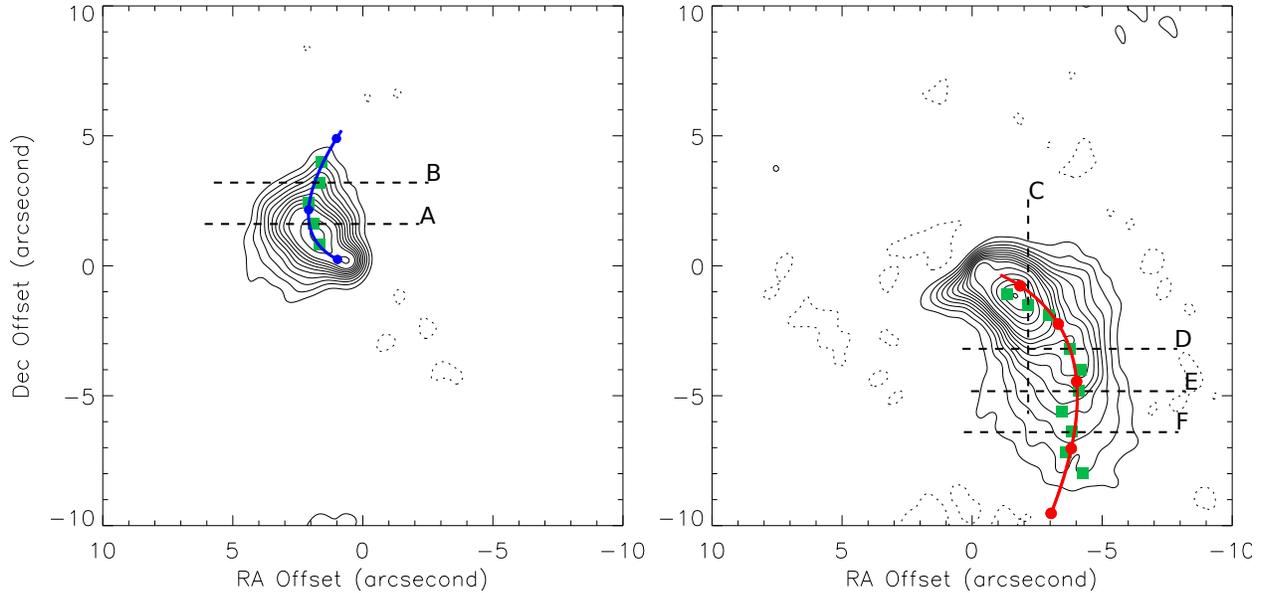}
\caption{Demonstration of our methods to extract spectra along the protrusions and to derive the trajectories of the infalling flows. Contours show moment 0 maps of the blueshifted (left) and redshifted (right) medium-velocity C$^{18}$O (2--1) emission in L1489 IRS (same as Figure \ref{C18Omap} middle). We first extracted the transverse intensity profiles of the protrusions by cutting along the RA or Dec directions (e.g., black dashed lines), and measured the positions of the intensity peaks along the transverse directions of the protrusions (green filled squares). Then we extracted spectra from these peak positions. The spectra at the peak positions, A, B, C, D, E, and F, are shown in Figure \ref{spec}. We then computed model trajectories of the infalling motion toward the protostar, and fitted the trajectories to these peak positions, which delineate the curvatures of the protrusions. Red and blue curves show the best-fit trajectories of the redshifted and blueshifted infalling flows, respectively. Red filled circles on the red trajectory denote the positional offset of 2\arcsec, 4\arcsec, 6\arcsec, 8\arcsec, and 10\arcsec\ from the center, and blue filled circles on the blue trajectory 1\arcsec, 3\arcsec, and 5\arcsec.}\label{infall}
\end{figure}

\begin{figure}
\epsscale{1}
\plotone{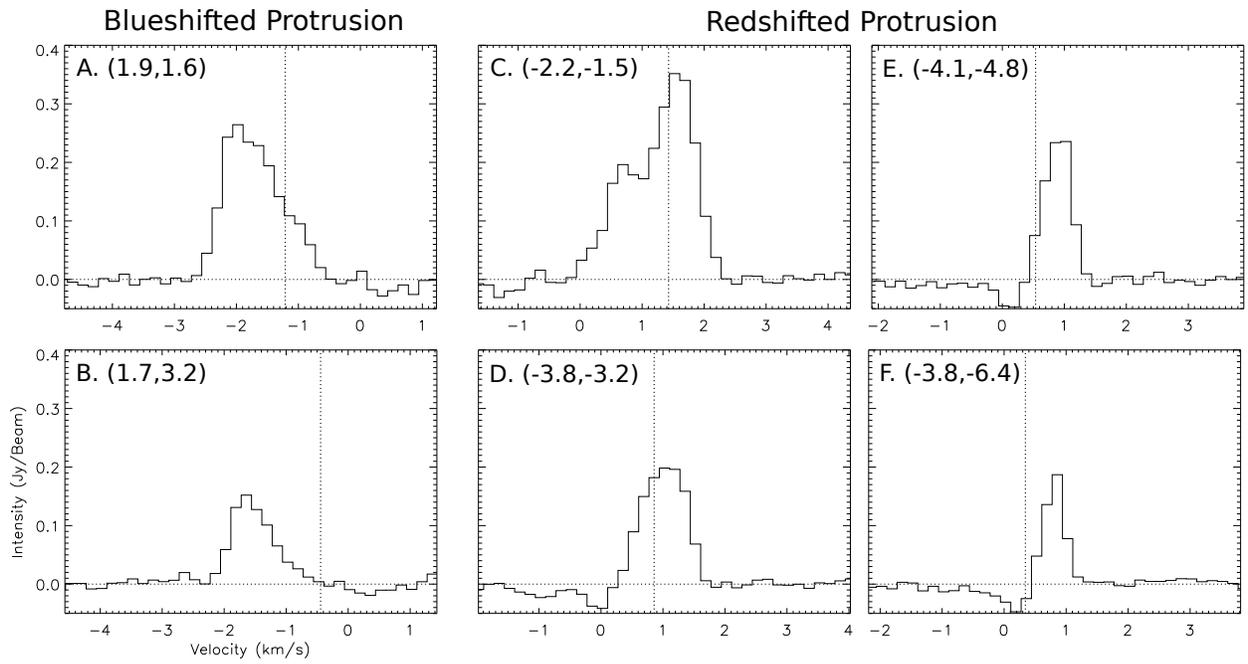}
\caption{Spectra at the peak positions along the blueshifted (left two panels) and redshifted (right four panels) protrusions (point A--F in Figure \ref{infall}). The RA and Dec offsets of the peak positions are shown at upper-left corner in each panel. Vertical dotted lines represent the expected velocities of the Keplerian rotation at these positions.}\label{spec}
\end{figure}

\begin{figure}
\epsscale{1}
\plotone{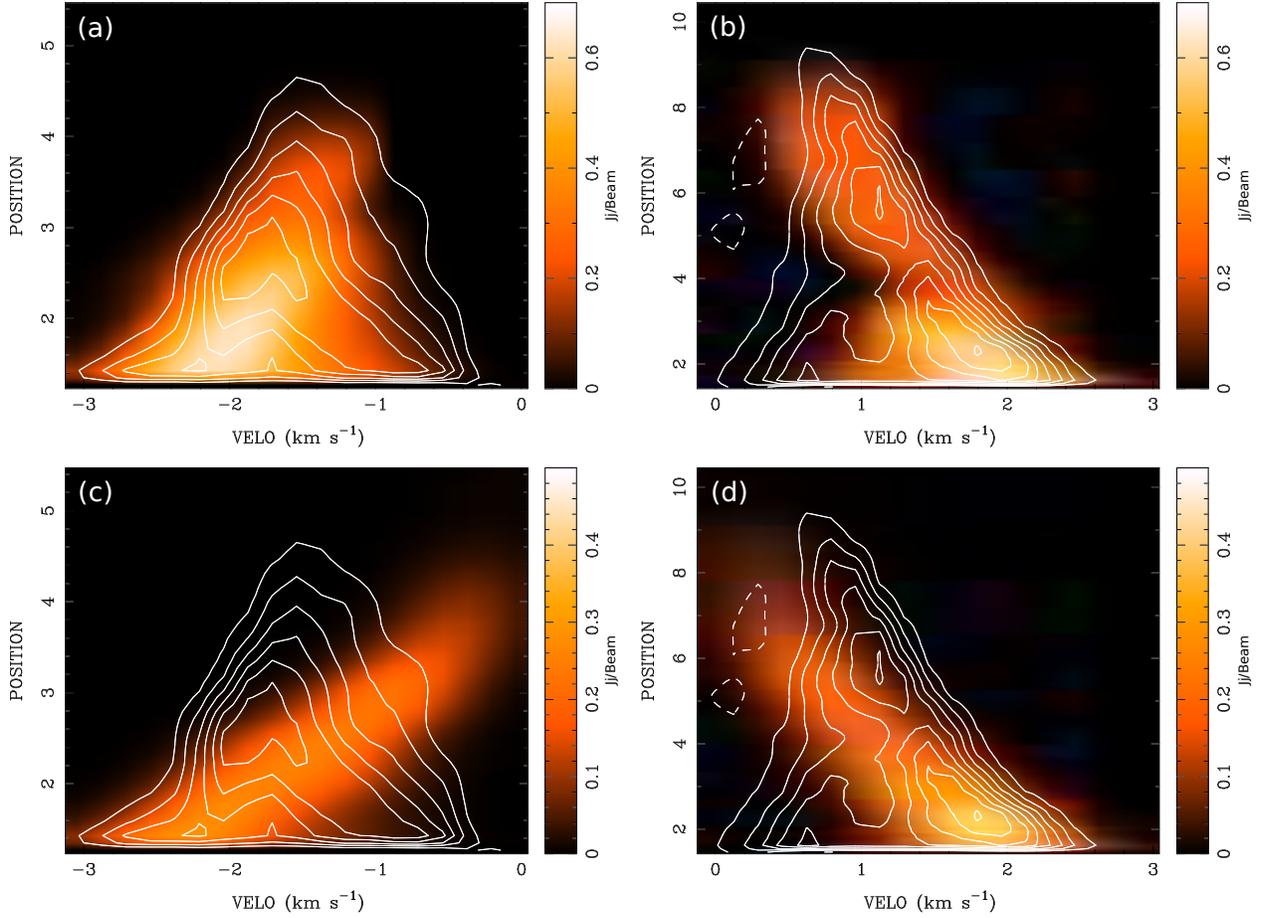}
\caption{P--V diagrams along the curvature of the blueshifted (left panels) and redshifted (right panels) protrusions (blue and red curves in Figure \ref{infall}), overlaid on the model P--V diagrams. Contours and color scales show the observed and model P--V diagrams, respectively. (a) \& (b) present the model of the infalling flows. (c) \& (d) present the model of the Keplerian rotation. Contour levels are from 5$\sigma$ in steps of 5$\sigma$, where 1$\sigma$ is 8 mJy.}\label{modelmap}
\end{figure}

From the total integrate C$^{18}$O flux at the medium velocity (Figure \ref{C18Omap} middle), the mass of the infalling flows is estimated to be 4--7 $\times$ 10$^{-3}$ $M_\sun$, assuming the LTE condition, excitation temperature of the C$^{18}$O emission of 10--40 K, and the C$^{18}$O abundance of 3 $\times$ 10$^{-7}$ (Frerking et al.~1982). 
In reality, there must be a contribution from the disk emission to the total C$^{18}$O flux at the medium velocity, 
and thus this should be over-estimation of the true mass of the infalling flows. 
On the other hand, from the comparison with the single-dish C$^{18}$O (2--1) spectrum obtained with the IRAM 30-m telescope (Hogerheijde et al.~1998), 
the missing flux in the medium-velocity range is estimated to be 20\%--50\%, 
so the mass of the infalling flows can be underestimated by a factor of 2--5. 
Therefore, this estimate of the mass of the infalling flows should be regarded as just an order estimate.
The infalling time scale is estimated to be $\sim$10$^{4}$ yr based on the lengths and infalling velocities of the infalling flows (Equation \ref{rin} and \ref{vin}). 
Thus, the mass infalling rate onto the Keplerian disk in L1489 IRS is estimated to be $\sim$4--7 $\times$ 10$^{-7}$ $M_\sun$ yr$^{-1}$, 
which is comparable to that in L1551 NE ($\sim$9.6 $\times$10$^{-7}$ $M_\sun$; Takakuwa et al~2013).
In L1489 IRS, 
one third of the total 1.3 mm continuum flux integrated over the 1$\arcmin$ (8400 AU) region measured by single-dish observations is originated from the central disk detected with our ALMA observations,  
and the total mass of the remaining envelope is $\sim$0.02 $M_\sun$ (Motte \& Andr{\'e} 2001). 
At the current mass infalling rate, it takes $\sim$3--5 $\times$ 10$^{4}$ yr to accrete all the remaining envelope material onto the Keplerian disk, 
which is one order of magnitude shorter than the typical timescale of the Class I stage (a few $\times$ 10$^5$ yr; Hatchell et al.~2007; Enoch et al.~2009). 
Assuming the luminosity of L1489 IRS ($L_{\rm bol}$ = 3.7 $L_\sun$) is fully supplied by the gravitational energy released by the material accreted from the disk onto the protostar, 
the mass accretion rate from the disk onto the protostar ($\tbond L_{\rm bol}R_*/GM_*$) is estimated to be $\sim$2 $\times$ 10$^{-7}$ $M_\sun$ yr$^{-1}$ with a protostellar radius of 3 $R_\sun$, comparable to the mass infalling rate onto the disk. 
Therefore, the Keplerian disk around L1489 IRS may continue to receive mass from the surrounding material and grow,   
while the disk mass should not exceed the current disk mass + remaining envelope mass (0.027 $M_\sun$). 
The remaining envelope mass ($\sim$0.02 $M_\sun$) is negligible to the current protostellar mass ($\sim$1.6 $M_\sun$), 
so the protostar likely has reached its final mass. 
Assuming that the disk will accumulate all the remaining envelope mass, 
by the end of the accretion phase, the upper limit of the mass ratio of the disk and the protostar in L1489 IRS is 2\%,  
which is similar to the mean ratio of $\sim$1\% in T Tauri stars (Williams \& Cieza 2011). 

Finally, we discuss the nature of the ring-like SO emission.  
According to the model by Aikawa et al.~(2012), 
SO is one of the major S-bearing molecules in infalling protostellar envelopes, 
and can be desorbed to the gas phase when the dust is heated to $\sim$60 K.
There are three possible heating mechanisms in protostellar sources, interaction with outflow, protostellar radiation, and accretion shock (e.g., Spaans et al.~1995; Bachiller \& P\'erez Guti\'errez 1997; Wakelam et al.~2005; van Kempen et al.~2009; Visser et al.~2009; Sakai et al.~2014).
Figure \ref{SOchan} and \ref{SOmap} show that the primary SO emission peaks in the blueshifted and redshifted velocities are, respectively, located to the northeast and southwest of the protostar at a radius of $\sim$300 AU from the central protostar, 
and the SO emission is elongated perpendicularly to the outflow direction. 
These results imply that the main SO components are unlikely to be associated with the outflow or protostellar radiation, 
except for the SO emission at low velocities ($V$ = $-0.8$ to 0.7 km s$^{-1}$) showing elongation along the outflow direction. 
The SO emission peaks are located at the positions adjunct to the infalling flows seen in the medium-velocity C$^{18}$O emission (Figure \ref{C18Omap} and \ref{SOmap}), and the velocity structures of the SO emission can be reproduced with our model of the ring region in the flared Keplerian disk (Figure \ref{modelpv}). 
In the chemical models by Aikawa et al.~(2012), the abundance ratio of $^{12}$CO/SO is $\sim$10$^4$ after SO is desorbed to the gas phase due to accretion shock. 
In L1489 IRS, the column density of SO is estimated to be $\sim$4 $\times$ 10$^{13}$ cm$^2$ to $\sim$1 $\times$ 10$^{14}$ cm$^2$ from its peak brightness temperature of 3--7 K and line width of $\sim$1 km s$^{-1}$ with kinematics temperatures of 20--30 K using RADEX (Van der Tak et al. 2007), 
while our best-fit disk model suggests that the column density of C$^{18}$O is $\sim$9 $\times$ 10$^{15}$ cm$^2$ at the radius of the SO ring. 
On the assumption that the abundance ratio of $^{12}$CO/C$^{18}$O is 250--560 (Wilson \& Rood 1994), 
the abundance ratio of $^{12}$CO/SO in the SO ring is estimated to be $\sim$2--10 $\times$ 10$^4$, 
comparable to the expected ratio in the chemical models within an order of magnitude.
Thus, we consider that the accretion shock is the most natural explanation of the localized SO emission enhancements. 
The infalling velocity at a radius of 300 AU is $\sim$3 km s$^{-1}$. 
The C-type shock models by Kaufman \& Neufeld (1996a,b) show that with a preshocked density of 10$^4$ to 10$^7$ cm$^{-3}$ and a shock velocity of 3 km s$^{-1}$, 
the shock can raise the temperature to 30--100 K, which is on the order of the SO desorption temperature $\sim$60 K. 
C-type shock models tend to produce molecular emission along the cavity walls (e.g., Visser et al. 2012). 
Therefore, the observed SO emission likely arises from the disk surface where the C-shocks tend to occur. 
Observations with higher spatial resolutions are required to resolve the vertical structures of the disk and pinpoint the regions emitting the SO emission.
The observed SO line width of $\sim$1 km s$^{-1}$ (Figure \ref{pv} right) corresponds to the thermal line width at a temperature of $\sim$150 K, higher than the expected temperature enhancement by the shock. 
Hence, the observed SO line width most likely reflects the turbulent gas motions in the shocked region but not the enhancement of the thermal line width associated with the shock. 
More detailed physical and chemical models of accretion shocks and further observations to resolve and measure the physical conditions of the SO emission are need to study the nature of the SO emission in L1489 IRS.

\section{Summary}
We have made ALMA cycle 0 observations of a Class I protostar L1489 IRS in the 1.3 mm continuum emission and the $^{12}$CO (2--1), C$^{18}$O (2--1), and SO (5$_6$--4$_5$) lines, 
and constructed models of a Keplerian disk and infalling flows to thoroughly reproduce the observed features. 
The main results are summarized below.

\begin{enumerate}
\item{The $^{12}$CO emission shows blueshifted and redshifted V-shaped features to the south and north of the protostar, respectively. These V-shaped features most likely trace the walls of the outflow cavities. In addition to these outflow components, there are high-velocity ($>$5 km s$^{-1}$), compact ($\sim$0\farcs5--0\farcs6) blueshifted and redshifted $^{12}$CO emission to the east and west of the protostar, respectively. The C$^{18}$O emission is elongated along the northeast--southwest direction ($\sim$1400 AU), perpendicular to the outflow axis, and exhibits a velocity gradient along the major axis, similar to that in the high-velocity $^{12}$CO emission. In the $^{12}$CO and C$^{18}$O P--V diagrams along the major axis the features of spin-up rotation are clearly identified. We have constructed models of a geometrically-thin Keplerian disk and performed $\chi^2$ fitting to the C$^{18}$O velocity channel maps in the high-velocity regime, and successfully reproduced the high-velocity C$^{18}$O emission as a geometrically-thin Keplerian disk with the protostellar mass of 1.6 $M_\sun$ and the inclination angle of 66\degr. We have also added the disk flaring with the scale height of 25 AU at $r$ = 100 AU to the best-fit geometrically-thin disk model, and succeeded to also reproduce the lower-velocity C$^{18}$O emission component exhibiting a linear velocity gradient in the P--V diagram.}

\item{The 1.3 mm continuum emission consists of a point source with the flux of $\sim$7.6$\pm$0.5 mJy and an elongated Gaussian component with a deconvolved size of 3\farcs5 $\times$ 1\farcs1 (500 AU $\times$ 150 AU), a position angle of 69\degr, and a total flux of $\sim$42$\pm$3.7 mJy. The elongation and extent of the Gaussian component are consistent with those of the Keplerian disk identified in the high-velocity C$^{18}$O emission, and thus the dust component most likely traces part of the Keplerian disk. The disk mass is estimated to be 0.003--0.007 $M_\sun$ with a dust temperature of 50--20 K from the 1.3 mm continuum flux.}

\item{At lower velocities ($|V|$ $<$ 2.5 km s$^{-1}$), the C$^{18}$O emission also shows blueshifted and redshifted ``off-axis'' protrusions pointing to the north and the south at the northeastern and southwestern tips of the main elongated structure, respectively. The line-of-sight velocities of these protrusions are 0.2--1.0 km s$^{-1}$ higher than those expected from the Keplerian rotation derived from the main C$^{18}$O component, and hence these off-axis components are likely distinct from the Keplerian disk. One possibility of the nature of the non-Keplerian protrusions is that they trace infalling flows onto the Keplerian disk. We have constructed models of infalling flows with a conserved angular momentum and zero total (potential and kinematic) energy, which are not co-planar to the disk plane. We found that the model with the centrifugal radius of 300 AU and initial injection polar and azimuthal angles of 52$\degr$ and 74$\degr$ for the blueshifted protrusion and 124$\degr$ and 262$\degr$ for the redishifted protrusion can reproduce the curvature and kinematics of the protrusions. The intensity peaks located at the junctional points of the protrusions and the main elongated structure may reflect the enhancement of the gas density due to the accretion of the material onto the disk.}

\item{While the SO emission also shows an elongated feature along the northeast--southwest direction and a velocity gradient along the major axis as in the case of the C$^{18}$O emission, the SO emission does not show a clear sign of spin-up rotation but a linear velocity gradient in the P--V diagram. One interpretation of the SO results is that the SO emission is primarily originated from a ring region in the flared Keplerian disk, and does not trace the inner high-velocity part of the Keplerian disk. We have demonstrated that a ring with the radial range from 250 AU to 390 AU in the flared disk obtained from the C$^{18}$O model fitting can reproduce the observed linear velocity gradient and the emission structure in the SO emission. The SO radial range is coincident with those of the C$^{18}$O intensity peaks adjunct to the non-Keplerian protrusions, and thus the SO ring may trace the shocked region where the infalling material falls onto the Keplerian disk.}

\item{Our models of the infalling flows implies that in L1489 IRS the material falls from the north and the south toward the Keplerian disk anisotropically, different from the cases of the other younger Class 0 and I sources having axisymmetric infalling envelopes. This could be due to the fact that the extended protostellar envelope around L1489 IRS has been dissipated and become asymmetric, and that only the remnants of the envelope material fall toward the central disk. The specific angular momentum of the infalling flows ($\sim$2.5 $\times$10$^{-3}$ km s$^{-1}$ pc) is comparable to that at the outer part of the Keplerian disk ($\sim$4.8 $\times$ 10$^{-3}$ km s$^{-1}$ pc). The mass infalling rate onto the Keplerian disk is estimated to be $\sim$4--7 $\times$ 10$^{-7}$ $M_\sun$ yr$^{-1}$, and the remaining envelope around L1489 IRS has a mass of 0.02 $M_\sun$. Therefore, the Keplerian disk around L1489 IRS may continue to grow in mass and size by accumulating mass and angular momenta from the infalling material.}
\end{enumerate}

\acknowledgments
This paper makes use of the following ALMA data: ADS/JAO.ALMA\#2011.0.00210.S. ALMA is a partnership of ESO (representing its member states), NSF (USA) and NINS (Japan), together with NRC (Canada) and NSC and ASIAA (Taiwan), in cooperation with the Republic of Chile. The Joint ALMA Observatory is operated by ESO, AUI/NRAO and NAOJ. We thank all the ALMA staff supporting this work. S.T. acknowledges a grant from the National Science Council of Taiwan (NSC 102-2119-M-001-012-MY3) in support of this work. K.Tomida is supported by Japan Society for the Promotion of Science (JSPS) Research Fellowship for Young Scientists.

\begin{figure}
\epsscale{0.5}
\plotone{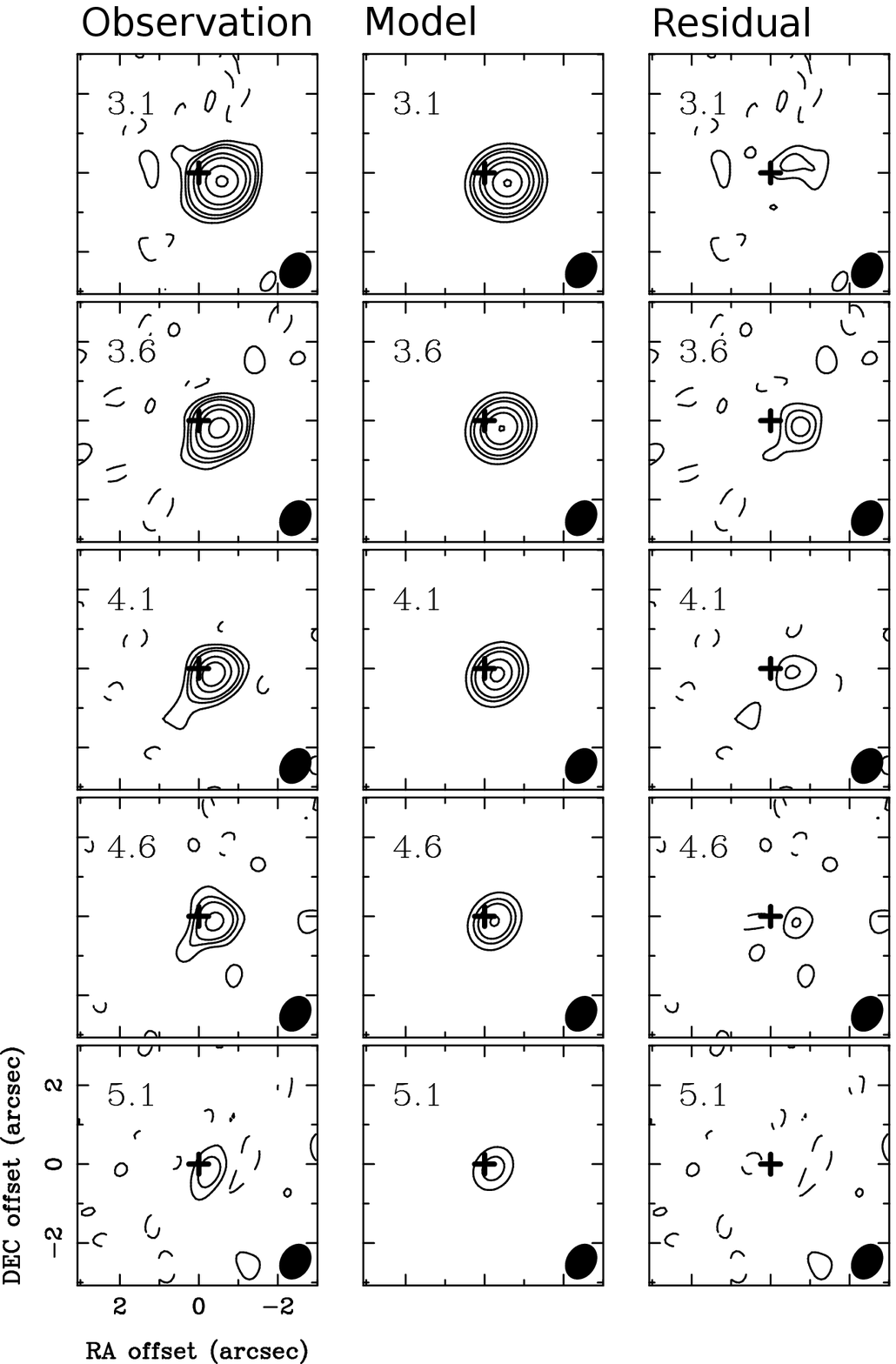} \\
{\bf Figure \ref{diskfit}.} (Continued)
\end{figure}

\clearpage

\begin{deluxetable}{ll}
\tablewidth{0pt}
\tablecaption{Best-Fit Physical Parameters of the Keplerian disk}
\tablehead{ \colhead{Parameter} & \colhead{Best-Fit Value} }
\startdata
{\smallskip}
Center of Mass ($\alpha_0$, $\delta_0$) & $4^{h}04^{m}43\fs07$$\pm$0$\farcs$2 \\ 
{\smallskip}
& $26\arcdeg18\arcmin56\farcs3$$\pm$0$\farcs$2 \\
{\smallskip}
Protostellar Mass ($M_*$) & 1.6$\pm$0.5 $M_\sun$ \\
{\smallskip}
Inclination ($i$) & 66$\pm$10$\degr$ \\
{\smallskip}
Position Angle ($\psi$) & 68$\pm$18$\degr$\\
{\smallskip}
Temperature at a Radius of 100 AU$\tablenotemark{a}$ ($T_0$) & 44 K\\
{\smallskip}
Power-law Index of the Temperature Profile$\tablenotemark{a}$ ($q$) & $-0.3$ \\
{\smallskip}
C$^{18}$O Surface Number Density at a Radius of 100 AU ($\Sigma_0$) & $6^{-3}_{+7} \times 10^{16}$ cm$^{-3}$\\ 
{\smallskip}
Power-law Index of the Surface Density Profile$\tablenotemark{a}$ ($p$) & $-1.7$ \\
{\smallskip}
Systemic Velocity ($V_{\rm  sys}$) & 7.3$\pm$0.1 km s$^{-1}$ \\
{\smallskip}
Turbulent dispersion$\tablenotemark{a}$ ($v_{\rm turb}$) & 0.3 km s$^{-1}$ \\
\enddata
\tablecomments{The distance of 140 pc to L1489 IRS is adopted.}
\label{fitting}
\tablenotetext{a}{The parameter is poorly constrained with our model fitting. See Section \ref{disk} for details.}
\end{deluxetable}

\end{document}